\documentclass[twocolumn]{revtex4-2}

\usepackage{mathtools,needspace,enumitem,etoolbox,physics,microtype,afterpage,bigints,tabularx,xspace}
\usepackage{graphicx, dcolumn, bm, lipsum}
\usepackage[commentmarkup=footnote]{changes}
\definecolor{linkcolor}{rgb}{0.0,0.3,0.5}
\usepackage[colorlinks = true,
            linkcolor = linkcolor,
            urlcolor  = linkcolor,
            citecolor = linkcolor,
            anchorcolor = linkcolor]{hyperref}
\usepackage{orcidlink}
\usepackage[normalem]{ulem}
\usepackage{pifont}
\usepackage{tabularx}
\usepackage[margin=1in]{geometry}
\usepackage{array}
\usepackage{subcaption}
\usepackage{float}
\usepackage{breqn}
\usepackage{booktabs}
\usepackage{multirow}

\allowdisplaybreaks[1]

\def\2pif{2\pi f_{\mathrm{obs}}}

\def\dto{dt_{\mathrm{obs}}}

\usepackage{array}

\usepackage{acronym}

\begin{document}

\title{Gravitational-wave detection of LISA massive black hole hierarchical triples via line-of-sight acceleration}

\newcommand{\tou}{\affiliation{Laboratoire des 2 infinis - Toulouse (L2IT), Université de Toulouse, CNRS/IN2P3, Toulouse, France}}

\author{Zheng Wu~\orcidlink{0000-0002-0032-5257}}
\email{zheng.wu@l2it.in2p3.fr}\tou
\author{Manuel Piarulli~\orcidlink{0009-0009-4099-9166}}
\email{manuel.piarulli@l2it.in2p3.fr}\tou

\author{Sylvain Marsat~\orcidlink{0000-0001-9449-1071}}
\email{sylvain.marsat@l2it.in2p3.fr}\tou
\author{Nicola Tamanini~\orcidlink{0000-0001-8760-5421}}
\email{nicola.tamanini@l2it.in2p3.fr}\tou

\date{\today}

\begin{abstract}
We consider massive black hole (MBH) binary signals observed by the future space-borne gravitational wave detector LISA, and investigate binaries whose center-of-mass undergoes a small acceleration, assumed constant during detection, such as when it orbits a distant tertiary mass.
The line-of-sight acceleration (LOSA) is incorporated as a new parameter that perturbs the \texttt{IMRPhenomXHM} waveform model.
The LOSA most significantly modifies the wave phase in the low-frequency portion of the signal, so lower-mass MBH binaries ($10^4$--$10^5\, M_\odot$) with extended inspiral in the LISA band are the most effective targets for detecting this effect.
Our Bayesian analysis reveals that LOSA measurement precision improves steadily for longer observational times in such low-mass systems, while high-mass systems ($10^7\, M_\odot$) are insensitive to the effect.
We also show that neglecting a sufficiently strong LOSA could induce systematic biases in other waveform parameters, in particular the chirp mass and the mass ratio.
Finally, we assess the validity of the constant LOSA approximation in realistic astrophysical scenarios and map the LOSA detectability to the mass-separation space of a possible nearby third body.
The LOSA induced by a $10^8\, M_\odot$ MBH circumbinary companion at separations up to 1 pc could, for example, be detected in the signal of $10^4$--$10^5\, M_\odot$ MBH binaries.
Our results establish the LOSA effect as a novel probe of MBH binary environments and as a possible mechanism to obtain information on MBH formation channels.
\end{abstract}

\maketitle

\section{Introduction}

Massive black hole binaries (MBHBs), consisting of two massive black holes (MBHs) with masses ranging from $\sim10^4$ to $10^7 M_\odot$, are one of the primary gravitational wave (GW) source targets of the \textit{Laser Interferometer Space Antenna} (LISA) \cite{LISA:redbook}, as well as of similar proposed GW observatories such as Taiji and TianQin \cite{TianQin:2015, TianQin:2020, Taiji:2021}.
The mechanisms behind the formation and evolution of MBHBs are still largely unknown, mainly due to the lack of observations with standard electromagnetic (EM) probes.
They are nevertheless believed to form as a consequence of galactic mergers.
MBHs are in fact ubiquitously found at the center of almost every galaxy, suggesting that after each galactic merger they sink to the galactic center of the newly formed galaxy through dynamical interactions with gas, stars, and eventually GW emission before ultimately coalescing.
Their detection and characterization with LISA will provide unprecedented insights into MBH growth, galaxy evolution, cosmology and strong-field gravity \cite{LISAastro:2022, LISACosmo:2022, LISAwaveform:2023}.

The coalescence pathways of MBHBs is governed by complex processes, including dynamical friction \cite{Vicente:2022ivh, Dosopoulou:2016hbg}, stellar scattering \cite{Baruteau:2010sm, Syer1991, Quinlan:1996vp, Dotti:2006ef}, and gas-driven migration \cite{Shi:2024skj, Dotti:2006ef}, yet the efficiency with which they overcome the so-called ``final parsec problem''---a potential stalling of the MBHB orbit at separations of $\sim$1 pc---remains uncertain \cite{LISAastro:2022}.
A possible solution to this problem involves triple interactions with another MBH, which may induce coalescence through the Kozai-Lidov mechanism or chaotic three-body dynamics \cite{Bonetti:2016eif, Bonetti:2017dan, Bonetti:2017lnj, Bonetti:2018tpf}.
In this scenario, the host galaxy of a stalled MBHB eventually merges with another galaxy which provides the third MBH.
MBHB mergers triggered by triple interactions can account for up to $\sim 30\%$ of the total MBHB merger rate \cite{Bonetti:2018tpf}.

Possible signatures on the GW signal imprinted by the third body can thus be detectable by LISA.

One of the main effects due to triple interactions is the imprint of a line-of-sight acceleration (LOSA) on the MBHB center of mass. 
This induces a time-dependent Doppler shift on the gravitational waveform, which effectively represents a $-4$PN additional de-phasing \cite{2017PhRvD..95d4029B, Tamanini:2019usx}.
The acceleration, arising from the gravitational pull of the third body on the binary, can significantly impact the detector frame chirp mass and luminosity distance if not properly measured.

No signals affected by acceleration have been so far identified in detections reported by the LIGO-Virgo-KAGRA (LVK) Collaboration \cite{Vijaykumar:2023tjg, Tiwari:2025aec, lonelyLVK, Lazarow:2024detectingaccelerating}.
LVK detectors are, however, only sensitive to the latest cycles of a compact binary merger.
Future detectors, on the other hand, will observe at least some GW signals, in particular low-mass ($\lesssim 10^5 M_\odot$) MBHBs, Extreme mass-ratio inspirals (EMRIs), and stellar-mass compact binary coalescences, starting from their deep inspiral phase (up to years before merger for space-borne detectors).
The LOSA effect imprints itself strongly during the early inspiral phase, making this stage particularly critical for its measurement. Future gravitational-wave observatories such as LISA, with their sensitivity to low-frequency signals, will therefore have a significantly better chance of detecting and characterising LOSA \cite{2010PhRvD..81f4018Y, Tamanini:2019usx} (though too far from merger the signal becomes almost monochromatic and the effect of the LOSA becomes degenerate with other waveform parameters \cite{2023PhRvD.107f4073B}).

In particular, stellar-mass binary black holes in dense environments are expected to exhibit detectable LOSA signatures for both gound- and space-based planned observatories \cite{Randall:2018lnh, Tiwari:2023cpa, Vijaykumar:2023tjg}.
EMRI could also carry phase drifts detectable by space-borne detectors induced by the acceleration of a nearby supermassive black hole (within $\sim$0.1–1 pc) \cite{Yunes:2010}.
LISA is also expected to observe a large number (tens of thousands) of binary white dwarfs within the Milky Way, which may as well be affected by the LOSA induced by a circumbinary companion \cite{Inayoshi:2017hgw, Tamanini:2018cqb, Danielski:2019rvt, Katz:2022izt}, or even by the much weaker gravitational potential of the Milky Way \cite{Ebadi:2024oaq}. 

Regarding MBHBs, the LOSA frequency drift in the waveform may be relevant if a sufficiently heavy and sufficiently close MBH lies close to the binary.
MBHBs are moreover expected to be recorded at a high signal-to-noise ratio (SNR) by LISA (up to thousands vs few tens for LVK sources), hinting at the possibility that even for signals close to merger with a low number of cycles the LOSA effect may not be negligible.
Furthermore, LISA will observe many MBHB mergers at high redshift, where galaxy mergers were more frequent and thus MBH triples more probable.

Our primary objective in this study is to find the detectable limits of the LOSA of MBHB systems detected by LISA, and consequently estimate the probability of detecting a third MBH depending on its mass and separation from the MBHB emitting GWs.
As shown below, we demonstrate that for low-mass ($\lesssim 10^5 M_\odot$) MBHB mergers, LISA will be able to detect the LOSA induced by MBHs above $10^8 M_\odot$ at separations $\lesssim 1$ pc, where we expect incoming MBHs to stall and accumulate (the last parsec problem) and where we expect to find third-body MBHs participating in chaotic triple interactions \cite{Bonetti:2016eif}.
In addition, we investigate how the LOSA may affect and bias the measurement of other parameters in the waveform if not properly taken into account.
We also perform analyses at different times-to-coalescence to show how the length of the analysed MBHB signal affects the constraining power on the LOSA.
Overall, our results highlight the importance of considering the effects of LOSA in the analyses of LISA MBHB systems, especially at their low-mass end ($\lesssim 10^5 M_\odot$). For these low-mass systems, although it carries a limited SNR, the long, early inspiral phase is found to be crucial for improving the measurement precision of the LOSA effect.

In addition, we explore the regime when the acceleration induced by the third body varies appreciably over the observation due to, for instance, a short outer orbital period. This could make the constant-LOSA approximation inadequate. To address this, we extend our waveform model to include the next-order correction, namely the line-of-sight jerk. We further quantify the regime of validity of the constant-acceleration assumption through injection–recovery tests, using the likelihood gain from adding the jerk as the criterion for its applicability.

The structure of the paper is as follows.
In Sec.~\ref{sec:parametrized_framework}, we introduced the parametric LOSA estimation method, and in Sec.~\ref{derivation LOSA} we  derive how the LOSA modifies the GW waveform through two different mathematical derivations. In Sec.~\ref{subsec:validity_inro}, the beyond-acceleration correction (jerk) is derived.
In Sec.~\ref{data}, we introduce some basic concepts of GW data analysis and describe the waveform model and the parameters used to simulate GW signals, and the methodology we employ to estimate these parameters. In Sec.~\ref{sec:results}, we present the results of our MCMC (Markov chain Monte Carlo) analysis for both specific MBBH systems and multiple randomly simulated MBBH populations. In Sec.~\ref{subsec:duration}, we show the influence of the observation period on the uncertainty of the LOSA measurement. In Sec.~\ref{subsec:otherparas}, we investigate how the inclusion of the LOSA parameter in the waveform model affects the uncertainty of the other parameters. In Sec.~\ref{bias}, we quantify the bias induced by neglecting the LOSA. In Sec.~\ref{subsec:validity_constant}, we assess the limit of validity of the constant LOSA approximation by comparing with waveforms that include the jerk. In Sec.~\ref{MCMC}, we translate the results obtained from our MCMC analyses into requirements for the third MBH to generate detectable LOSAs.

In what follows we adopt a flat $\Lambda$CDM cosmology with $H_0=67.74\, {\rm km} / ({\rm Mpc}\times  {\rm {s}})$ and $\Omega_m=0.3075$ \cite{Planck:2015fie}.
The typical MBHB source used in our analyses is placed at $z = 1$ and assumes a mass ratio $q=m_1/m_2=3$, unless otherwise specified.

\section{Theoretical framework}
\label{theory}

In this section we outline the theoretical framework used in our analysis.
We first introduce the general approach of parametrized inspiral phase deviations.
Then we derive the effect of LOSA on the phase of the waveform, showing, via two independent calculations, that at first order it formally corresponds to a $-$4PN effective deviation in the inspiral phase.
Finally, we discuss the regime of validity of a constant LOSA for realistic third-body companions on a Keplerian orbit, and we derive the jerk contribution to the phasing.

\subsection{Parametrized inspiral phase deviations}
\label{sec:parametrized_framework}

In this work, we probe the dynamics of triple systems by employing the parametrized framework recently developed in Ref.~\cite{Piarulli:2025rvr}, which adapts the Flexible Theory-Independent (FTI) pipeline~\cite{FTI} to the context of LISA MBHB observations. 
While originally constructed to perform null tests of GR, the mathematical framework is equally well-suited for constraining macroscopic astrophysical effects.

We firstly focus on the inspiral phase of quasi-circular, non-precessing binaries, where the dynamics are accurately described by the post-Newtonian (PN) approximation~\cite{2024LRR....27....4B}. 
The frequency-domain gravitational waveform for a given $(l,m)$-mode can be expressed as $\tilde{h}_{\ell m}(f) = A_{\ell m}(f) \exp[i \psi_{\ell m}(f)]$ \cite{1999PhRvD..59l4016D, 2000PhRvD..61j4015T, 2009PhRvD..80h4043B}. 
The baseline general relativity (GR) phase is perturbatively expanded in powers of the PN expansion parameter $v=\left({2\pi M f/m}\right)^{1/3}$ as: 

\begin{multline}
    \Psi^{\mathrm{GR}}_{lm}(f) = \frac{3}{128\eta v^5}\frac{-m}{2} \\
    \left[ \sum_{n=0}^{7} \psi^{\mathrm{PN}}_n v^n + \sum_{n=5}^{6} \psi^{\mathrm{PN}}_{n(\ell)} v^n \log v \right] \\
    \quad + [\text{const}] + [\text{linear term}],
\label{eq:psi_GR}
\end{multline}
where $\psi^{\mathrm{PN}}_n$ and $\psi^{\mathrm{PN}}_{n(\ell)}$ are the $(n/2)$-PN coefficients dependent on the binary's intrinsic parameters~\cite{2009PhRvD..80h4043B, 2024LRR....27....4B}, $M$ is the total mass of the binary and $\eta=m_1m_2/M^2$.
To describe beyond-GR effects, the FTI framework introduces parametric deviations to each PN coefficient in GR of the form
\begin{equation}
    \psi_i \longrightarrow (1 + \delta\hat\varphi_i)\psi^{\mathrm{PN}}_i,
\label{eq:GR_to_FTI}
\end{equation}
with $\delta\hat\varphi_i$ as the deviation parameter, resulting in a global parameterized phase shift $\delta\Psi_{lm}(f)$. The framework also includes a tapering of the deviation when reaching the merger part of the signal, which is left unaffected; see Ref.~\cite{Piarulli:2025rvr} for further details.

The key novelty of the present work is the extension of this framework to test the environmental effect driven by LOSA, entering at $n = -8$~\cite{2014PhRvD..89j4059B, 2021PhRvL.126j1105T, 2017PhRvD..95d4029B, Tamanini:2019usx}. By contrast with Eq.~\eqref{eq:GR_to_FTI}, since GR does not feature such an effective PN order, the modification is written as a new term in the phasing, not a relative deviation (as was done in ~\cite{Piarulli:2025rvr} e.g.~for dipolar radiation with $n=-2$).
By integrating a new, dedicated parameter for this $-4$PN effect, the total modified phase used in our analysis becomes:
\begin{equation}
    \Psi_{lm}(f) = \Psi^{\mathrm{GR}}_{lm}(f) + \delta\Psi^{\mathrm{acc}}_{lm}(f) \,,
\end{equation}
where the phase correction includes only the $n=-8$ PN term.

In the following section, we provide a formal derivation for the $-4$PN term arising from the LOSA and establish the kinematic limits of its application.

\subsection{Derivation of the line-of-sight-acceleration correction}\label{derivation LOSA}

Here we present two equivalent ways to derive the LOSA modification on the phase of the GW waveform.
The first derivation is similar to the ones already reported in the literature \cite{2017PhRvD..95d4029B, Tamanini:2019usx}, and only the main details will be provided here.
The second derivation is instead more general and intuitive. It is based on applying a time delay induced by the LOSA directly in the waveform, similarly to the approach used in more general perturbations due to a third body \cite{2021PhRvL.126j1105T, Sberna:2022qbn}.
We provide the explicit expressions  at  1PN  for illustrative purposes, and more extensive expressions are given in App.~\ref{app:full_expression}.

Let us start with the first derivation.
To derive the effect of the LOSA on the GW waveform, we start from the energy $E$ and GW luminosity $\mathcal{L}_{\rm GW}$ of a compact binary up to 1PN order:
\begin{equation}
    E=-\frac{1}{2}M\eta v^{2}\left[1-\left(\frac{3}{4}+\frac{1}{12}\eta\right)v^2\right] \,,
    \label{eq:orbital_energy}
\end{equation}
\begin{equation}
    \mathcal{L}_{\rm GW}=\frac{32}{5}\eta^2 v^{10}\left[1-\left(\frac{1247}{336}+\frac{35}{12}\eta\right)v^{2}\right] \,.
    \label{eq:gw_luminosity}
\end{equation}

We define $z_{p}=\Gamma \times t_{\rm obs}$ to be the redshift correction induced by the LOSA of the binary, where $\Gamma=a^{\parallel} / c$, $a^\parallel$ is the LOSA, and $t_{\rm obs}$ is the time at the observer.
Ignoring the cosmological redshift, which for the purpose of our investigation can always be considered constant and thus absorbed in a redefinition of the masses and distance (see \cite{2017PhRvD..95d4029B, Tamanini:2019usx} for the inclusion of varying cosmological redshift), the time in the observer frame $\dto$ can be related to the time in the source frame $t$ by $dt=\dto/(1+z_{p})$, which implies an inverse relation for the corresponding frequencies $f_{\rm src}=f_{\rm obs}(1+z_{p})$.

Following the usual approach to derive the emission of GWs from a slowly evolving binary, we equate the time derivative of the orbital energy \eqref{eq:orbital_energy} to the GW luminosity \eqref{eq:gw_luminosity}
\begin{equation}
    \mathcal{L}_{\rm GW}=-\frac{dE}{dt}=-\frac{dE}{\dto}(1+z_{p}) \,,
\end{equation}
which, after some manipulation (see \cite{2017PhRvD..95d4029B, Tamanini:2019usx}) leads to the following expression for the phase of the waveform in the Fourier domain
\begin{equation}
    \begin{aligned}
     \Psi_{lm}(f)&=2\pi ft(f)-2m\phi(f)-\frac{\pi}{4}\\&=2\pi ft_{c}-m\phi_{c}-\frac{\pi}{4}.\\&+\frac{3m}{128\eta}v^{-5}\left[1+\frac{20}{9}\left(\frac{743}{336}+\frac{11}{4}\eta\right)v^{2}\right.\\& +\left.\frac{25\Gamma M}{1536\eta}v^{-8}\left\{1+\left(\frac{743}{126}+\frac{22}{3}\eta\right)v^{2}\right\}\right]\,,
    \end{aligned}
    \label{phaseofv}
\end{equation}
where $m$ is the azimuthal order in the harmonic mode $(l,m)$, and $\phi(f)$ is the orbital phase of the binary system.

The second derivation we provide is based on the Fourier-domain perturbative approach of \cite{Marsat:2018oam}, specifically the leading-order local-in-frequency approximation. Under our assumption that the outer third-body orbit is much longer than the inner binary orbit,  we calculate the unperturbed waveform in the source frame and subsequently apply the time-dependent delay due to the line-of-sight acceleration from the third body. The observed waveform in the frequency domain is approximated by a multiplicative transfer function,
\begin{equation}
    \begin{aligned}\tilde{h}_{\mathrm{obs}}(f)&\simeq\int dt\ h(t+d_D(t))\exp\left[2\pi ift\right]\\&\simeq\tilde{h}_{\rm src}(f)\exp\left[-2\pi ifd_{D}(t_{f})\right]\\&=A(f)\exp\left[-i\varphi(f)\right]\exp\left[-2\pi ifd_{D}(t_{f})\right]\,,\end{aligned}
    \label{SPAeq}
\end{equation}
where $\tilde{h}_{\rm src}(f)=A(f)\exp\left[-i\varphi(f)\right]$ is the vacuum GW signal at the source. Some quantitative conditions, such as $\dot{d}_{D(t)} \ll 1$ (which will be detailed at the end of this section), are used to justify the approximation in the second line. In our simple case, the delay caused by the constant LOSA is
\begin{equation}
    d_D(t_f)=\frac{1}{2}\frac{a^{\parallel}t_{f}^2}{c}=\frac{1}{2}\Gamma t_{f}^2,
    \label{delayeq}
\end{equation}
assuming that the time $t_f$ is measured relative to a point of reference where the base redshift is defined (typically, the merger), that is to say that $t_f = 0$ corresponds to the time of coalescence.
The time-frequency correspondence follows directly from the Fourier-domain phase derivative \cite{Marsat:2018oam},
\begin{equation}
    t_f = -\frac{1}{2\pi}\frac{d\Psi}{df} \,.
    \label{tfinitial}
\end{equation}

Notably, unlike the Stationary Phase Approximation (SPA), this approach does not require the time-domain adiabatic conditions $|\dot{\omega}/\omega^2|\, \ll 1$. Instead, it applies to any inspiral--merger--ringdown (IMR) waveform whose Fourier-domain amplitude and phase remain smooth, and therefore extends naturally through the merger and ringdown phases~\cite{Marsat:2018oam}.

At one 1PN order, the Fourier-domain phase of the unperturbed, i.e.~not delayed, signal is 
\begin{equation}
\varphi(f)=-\frac{3}{128\eta}v^{-5}\left[1+\frac{20}{9}\left(\frac{743}{336}+\frac{11}{4}\eta\right)v^{2}\right]\,,
\label{PN1phi}
\end{equation}
Substituting Eq.~\eqref{PN1phi} into Eq.~\eqref{tfinitial} yields
\begin{align}
    t_f&=-\frac{5M}{256\eta}v^{-8}\left[1+\left(\frac{743}{252}+\frac{11}{3}\eta\right)v^2\right. \nonumber\\& \left.+\frac{65\Gamma M}{1536\eta}v^{-8}\left\{1+\frac{22}{13}\left(\frac{743}{252}+\frac{11}{3}\eta\right)v^{2}\right\}\right]
\label{toff2}
\end{align}
Substituting Eq.~\eqref{toff2} into Eq.~\eqref{delayeq} and Eq.~\eqref{SPAeq}, we obtain the same result of Eq.~\eqref{phaseofv}.
Such an alternative derivation is more intuitive and can also be applied to the case where the LOSA is not constant (see e.g.~\cite{Sberna:2022qbn,2021PhRvL.126j1105T, Pompili:2026tdf}).

The Fourier-domain perturbative approach of~\cite{Marsat:2018oam} is controlled by a set of dimensionless error measures $\epsilon$. For $\epsilon \ll 1$, the perturbative approach applies, and $\epsilon$ can be used as an estimate for the magnitude of
the effect. In our application, we define:
\begin{equation}
    G(f,t) = e^{-2i\pi f d(t)}
           = \exp\!\left[-i\pi f\Gamma t^{2}\right],
\end{equation}
with \(d(t)=\tfrac12\Gamma t^{2}\) and \(t=t_f-t_c\), measured relative to the time of coalescence, as introduced above. The error measures of~\cite{Marsat:2018oam} read (besides less important $\epsilon_{A_1}$, $\epsilon_{A_2}$ that we will ignore here)
\begin{align}
    \epsilon_{\Psi 2} &= \frac{1}{2} T_f^{2}
        \left|\frac{1}{G}\partial_t^{2} G\right|, &
    \epsilon_{d} &= \frac{1}{2\pi}
        \left|\frac{1}{G}\partial_{tf} G\right| ,
\end{align}
where \(T_f\) is defined from the Fourier-domain phase
\(\Psi(f)\) of~\cite{Marsat:2018oam} as
\begin{equation}
    T_f^{2} \equiv \frac{1}{4\pi^{2}}
        \left|\frac{d^{2}\Psi}{df^{2}}\right| .
    \label{eq:Tf-def}
\end{equation}
Substituting \(G(f,t)\) gives
\begin{align}
    \frac{1}{G}\partial_{tt}^{2} G
        &= -2i\pi f\Gamma - 4\pi^{2} f^{2}\Gamma^{2} t^{2}, \\
    \frac{1}{G}\partial_{tf} G
        &= -2i\pi\Gamma t - 2\pi^{2} f\Gamma^{2} t^{3},
\end{align}
so that
\begin{align}
    \epsilon_{\Psi 2}
        &= \frac{1}{2} T_f^{2}
           \left|2\pi f\Gamma + 4\pi^{2} f^{2}\Gamma^{2} t_f^{2}\right|, \\
    \epsilon_{d}
        &= \left|\Gamma t_f + \pi f\Gamma^{2} t_f^{3}\right| .
\end{align}

As an illustrative estimate we take \(q=3\),
$\Gamma=10^{-10} s^{-1}$ and $f=f_{10 \; \mathrm{yr}}$, the frequency corresponding to ten years before the merger. Evaluating
Eq.~\eqref{eq:Tf-def} together with Eq.~\eqref{toff2} for total masses
\(M=10^{4}\)--\(10^{7}\), we obtain
\(\epsilon_{\Psi 2}\sim 10^{-7}\)--\(10^{-8}\) and
\(\epsilon_{d}\sim 10^{-9}\)--\(10^{-10}\); \(\epsilon_{\Psi 2}\) peaks at
\(\approx10^{-7}\) for \(M=10^{4}\)
while \(\epsilon_{d}\) peaks at \(\approx10^{-9}\) for
\(M=10^{7}\). Both measures are thus well
below unity.

\subsection{Beyond the constant line-of-sight acceleration approximation}
\label{subsec:validity_inro}

In extreme regimes, when the outer orbital period is  short enough, the Doppler modulation makes the GW frequency evolution itself
non-monotonic, with alternating chirping and anti-chirping phases, so that the
time--frequency map becomes multivalued~\cite{Sberna:2022}. The usual frequency-domain
perturbative method, which assumes monotonic frequency evolution and a
single-valued time--frequency correspondence, is no longer applicable in such a regime. As a proxy for evaluating the validity of our approach, beyond computing the $\epsilon_{\Psi 2}$ and $\epsilon_d$ quantities, we can consider the magnitude of the first correction beyond a constant-acceleration trajectory, which is given by the jerk $j$ in a Taylor expansion, and compute the likelihood penalty that would be induced by ignoring this term.

The expression for the time delay caused by the non-constant-velocity movement of the CoM of the BBHs is as follows:
\begin{equation}
    d_D(t_f)=\frac{1}{c} \left(\frac{1}{2}a^{\parallel}t_{f}^2+\frac{1}{6}j^{\parallel}t_{f}^3 \right)\,,
    \label{delayetotal}
\end{equation}

where $j^\parallel$ is the line-of-sight \textit{jerk}, the time derivative of the LOSA.

When the line-of-sight jerk is included on top of the LOSA, repeating the steps of the derivation above, the phase of the GW is obtained as: 

\begin{multline}
     \Delta\Psi_{lm}(f)=2\pi f d_{D}(t_f)\\
     =\frac{3m}{128\eta}v^{-5}\left[1+\frac{20}{9}\left(\frac{743}{336}+\frac{11}{4}\eta\right)v^{2}\right.\\
     \quad+\frac{25\Gamma M}{1536\eta}v^{-8}\left\{1+\left(\frac{743}{126}+\frac{22}{3}\eta\right)v^{2}\right\}\\
     \quad+\left.\frac{125jM^{2}}{1179648c\eta^{2}}v^{-16}\left\{1+\left(\frac{743}{84}+11\eta\right)v^{2}\right\}\right]\,
    \label{phaseofvplusj}
\end{multline}

In Sec.~\ref{subsec:validity_constant}, we will use the template with constant LOSA to recover an injected signal with time-varying LOSA (non-zero jerk) in order to determine the regime of validity of our constant LOSA approximation.
In particular, we will use the $\log\mathcal{L}$ threshold to determine the boundary where the constant LOSA approximation holds.

\section{Data analysis setup}
\label{data}

We implement the LOSA effect within the \texttt{lisabeta} framework \cite{Marsat:2020rtl}, which provides a comprehensive LISA response accounting for both the detector's orbital motion and Time Delay Interferometry (TDI) \cite{Tinto:2004wu}. We employ the first-generation noise-orthogonal TDI variables $A$, $E$, and $T$ under the approximation of equal and static arm lengths  \cite{Tinto:2004wu, Hartwig:2022yqw}. The instrumental response is evaluated in the Fourier domain following \cite{Marsat:2020rtl}, with the key modification being the acceleration-induced phase shifts $\delta\Psi_{\ell m}$. These alter the time-of-frequency relation through $\delta t_{\ell m}(f) = (2\pi)^{-1}\, d\delta\Psi_{\ell m}/df$, computed via differentiation of a cubic interpolating spline. The resulting TDI signals take the sparse Fourier-domain form (representing data with a few hundred points in frequency space)
\begin{equation}
 \tilde{s}_k = \sum_{\ell m} T^k_{\ell m}(f)\tilde{h}_{\ell m}(f) \exp \bigl[i\delta\Psi_{\ell m}(f)\bigr] ,
 \label{eq:tdi_signal}
\end{equation}
where $T^k_{\ell m}$ are the mode-wise TDI transfer functions, $k$ is an index representing the TDI channel $A,E,T$, with time-dependent quantities evaluated at the modified time-of-frequency $t_{\ell m}(f) + \delta t_{\ell m}(f)$. We refer to \cite{Marsat:2020rtl} for further notation.

As our baseline waveform, we adopt the aligned-spin inspiral-merger-ringdown model \texttt{IMRPhenomXHM} \cite{Pratten:2020fqn, Garcia-Quiros:2020qpx}, which includes the spherical harmonic modes $(\ell,m) = (2, 2), (2, 1), (3, 3), (3, 2), (4, 4)$. To incorporate LOSA effects, we extend this model by introducing the acceleration-induced phase modifications expressed in Eq.~\eqref{phaseofv}. While more recent frameworks include precession and eccentricity 
\cite{Pratten:2020ceb, Planas:2025feq}, we restrict this study to spin-aligned, quasi-circular binaries.
A discussion on extensions to the most complex waveform models is reported in Sec.~\ref{sec:conclusions}.

The SNR for a single source is defined as
\begin{equation}
 \rho^2 = \sum_{k} (s_k|s_k)_k,
 \label{eq:snr}
\end{equation}
with $k \in \{A,E,T\}$ indexing the noise-orthogonal TDI channels. The inner product between two functions $a(t)$ and $b(t)$ in channel $k$ is
\begin{equation}
 (a|b)_k = 4\operatorname{Re} \int_0^\infty df \frac{\tilde{a}^*(f)\tilde{b}(f)}{S_{n,k}(f)},
 \label{eq:inner_product}
\end{equation}
where $S_{n,k}(f)$ is the power spectral density. Under the assumption of stationary, Gaussian, and mutually independent noise across channels, the log-likelihood for data streams $d_k$ given a signal model $s_k(\theta)$ is
\begin{equation}
 \ln \mathcal{L} = -\frac{1}{2} \sum_{k=A,E,T} \bigl(s_k(\theta)- d_k \big| s_k(\theta)- d_k\bigr)_k ,
 \label{eq:likelihood}
\end{equation}
where ${\theta}$ denotes the physical parameters, including the LOSA. The data consist of a true signal $s_k({\theta}_0)$ plus a noise realization $n_k$: $d_k = s_k({\theta}_0) + n_k$.

While the Fisher information matrix, $F_{ij} = \sum_k (\partial_i s_k | \partial_j s_k)_k$, offers a computationally inexpensive Gaussian estimate of parameter uncertainties, it is inadequate for capturing the complex correlations and non-Gaussian posterior structure --- particularly at negative PN orders such as the LOSA parameter.
A detailed discussion about the failure of the Fisher matrix approximation in our case is provided in Appendix \ref{app:fisher_low_freq}.
We therefore
conduct our primary inference through full Bayesian analysis, on which all our main results are based.

Using the \texttt{lisabeta} package \cite{Marsat:2020rtl},  we sample the posterior distribution
\begin{equation}
 p(\theta|d) = \frac{p(d|\theta) p(\theta)}{p(d)},
 \label{posterior}
\end{equation} 
where $p(\theta)$ encodes the parameter priors. As our goal is to characterize parameter estimation rather than perform model comparison, we treat the evidence $p(d)$ as a normalization constant.
We adopt the zero-noise approximation ($n_k = 0$), which is sufficient for studying parameter degeneracies and likelihood morphology. For computational efficiency, waveforms and inner products are evaluated on a non-uniform frequency grid: finely sampled at low frequencies where the long inspiral accumulates many phase cycles, and coarsely sampled at high frequencies where the short merger contributes fewer cycles. Posterior exploration is performed with the parallel-tempered ensemble sampler \texttt{ptemcee} \cite{Vousden:2016eeu}\cite{emcee}, enabling robust sampling of multi-modal distributions. The one-dimensional posterior for the LOSA parameter is finally obtained by marginalizing over all other parameters.

\section{Results}
\label{sec:results}

In this section, we present the results of our investigation focusing on how the duration of the signal, depending on both the total mass of the MBHB and the actual observational time of LISA, affects the detectability of the LOSA.
Subsequently, we present how the presence of a LOSA affects the measurement of other parameters of the waveform. We truncate the phase correction of LOSA at the next-to-leading order ($-3$ PN), since the resulting likelihood penalty relative to the full PN expansion is negligible across the astrophysical scenarios considered (see details in App.~\ref{app:-4PNvsfullPN}).

\subsection{Effect of signal duration on LOSA measurements}
\label{subsec:duration}

\begin{figure*}[!htb]
    \centering
    \includegraphics[width=\textwidth]{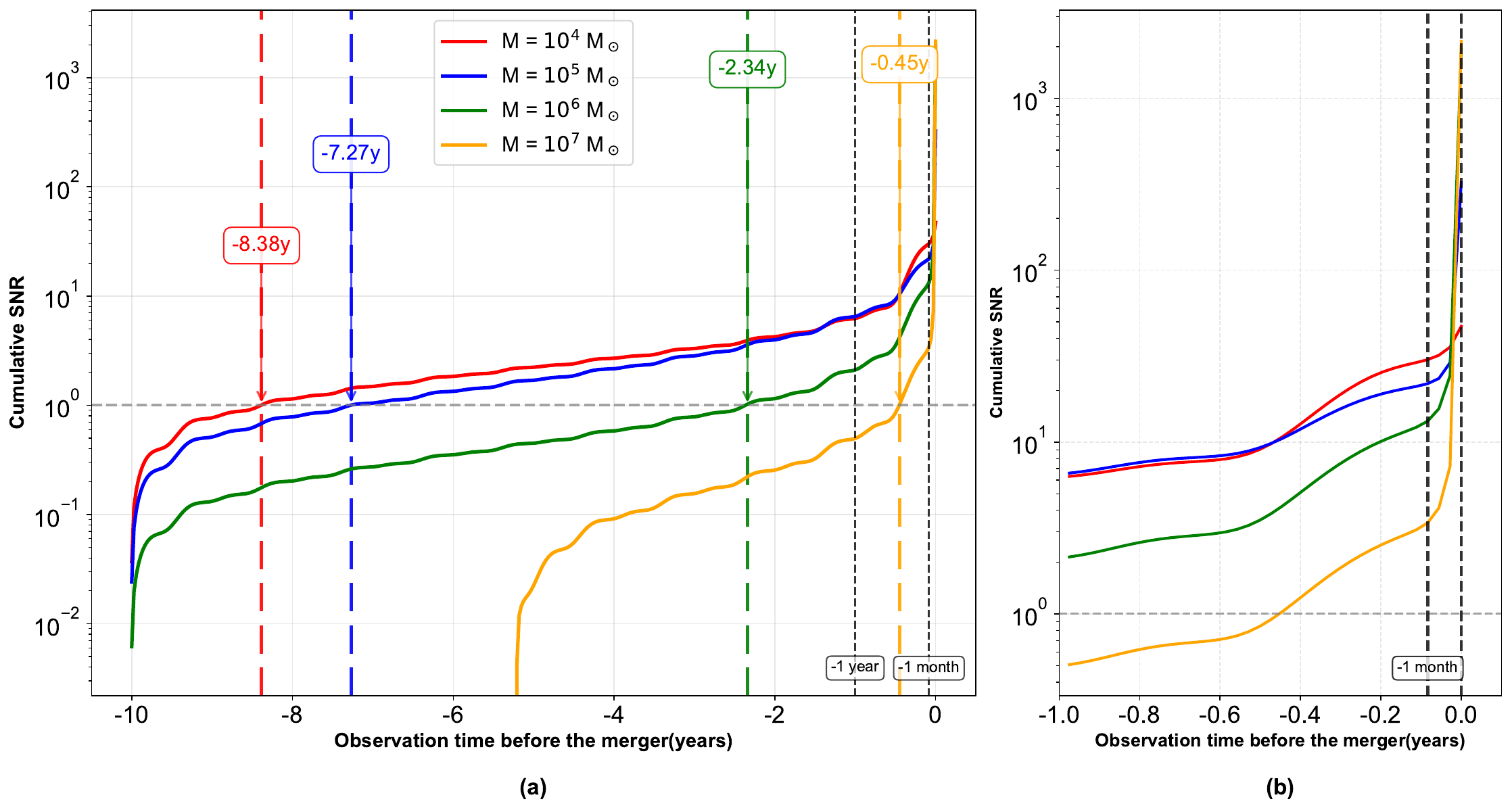} 
    \caption{\textbf{(a)} Cumulative (2,2) mode SNR vs time before merger for MBHBs at $z=1$ with total mass $M = 10^4$--$10^7\, M_{\odot}$. From the lower to the higher mass, the SNRs at merger are 47, 323, 1933, and 2149, respectively. The vertical colored dashed lines mark the SNR=1 crossing time.
    The remaining source parameters chosen for the injections are listed in Table~\ref{other paras} in App.~\ref{appendix paras}.
    }
    \label{fig:snroft}
\end{figure*}

Let us consider the likelihood Eq.~\eqref{eq:likelihood}, with the injection being a signal $s$ with zero LOSA $(a^\parallel = 0)$ (for one of the TDI channels and one of the harmonics of the signal, say the dominant $(2,2)$-mode), and the template being given for the same parameters except $a^\parallel \neq 0$. Since the effect of LOSA can be written mostly as a Fourier-domain phase correction, the per-log-frequency likelihood penalty between such an injection and template is bounded by the log-frequency density of $\rho^2$ by
\begin{equation}
\frac{d[(-\ln\mathcal{L})]}{d\ln f}
\simeq \frac{1}{2} |1-e^{-i\Delta\Psi}|^{2}\,\frac{d\rho^{2}}{d\ln f},
\label{eq:snr_likelihood_density}
\end{equation}
where $\Delta\Psi$ is the phase difference caused by acceleration, $s(a^\parallel) \simeq e^{-i \Delta \Psi} s(a^\parallel = 0)$. The presence of the acceleration also causes a small difference in the time-to-frequency map $t_f$, and a correspondingly small change in the instrumental response $\mathcal{T}(f, t_f)$, which we ignore here.

The differential relation shows that the logarithmic likelihood density and the squared SNR density are linked by a frequency-dependent factor bounded by $[0,2]$. The likelihood penalty is locally maximal, saturating the bound 2, when the two signals are in phase opposition with $\Delta \Psi(f) = \pi$. Over an interval where the phase difference is large and varies rapidly with frequency, effectively randomizing $\Delta \Psi$, the prefactor averages as $\langle 2 \sin^2 (\Delta \Psi /2)\rangle_{\Delta \Psi} = 1$. This bound also makes explicit the fact that parts of the frequency intervals contributing a negligible amount to the total SNR will also contribute little to the likelihood penalty, and are therefore uninformative. To set the scale of likelihood values, we note that, for a likelihood approximated as a $d$-dimensional Gaussian distribution, the log-likelihood ratio $-2 \ln (\mathcal{L} / \mathcal{L}_{\rm max})$ is distributed as a chi-squared with $d$ degrees of freedom, so that $\langle \ln (\mathcal{L} / \mathcal{L}_{\rm max}) \rangle = -d/2$.

According to Eq.~\eqref{phaseofv}, the low-frequency part in the inspiral stage of the signal contributes the majority of the phase differences caused by the LOSA.
The duration of a MBHB signal from its entry into the LISA frequency band to its merger depends on its total mass $M$ (mainly through the chirp mass).
Consequently, the rate of SNR accumulation varies significantly depending on the total mass of the binary black hole system.
For systems with lower masses, their GW signal has a longer duration within the LISA frequency band, and the early inspiral stage contributes more to the SNR, whereas higher-mass binaries enter the LISA frequency band relatively late and thus present a lower SNR accumulation during the early inspiral.
In the following, we investigate this relationship quantitatively and demonstrate how the observation duration before merger affects the lower bounds on the detectable LOSA of MBHB systems.

Fig.~\ref{fig:snroft} shows a comparison of the cumulative SNR of MBHBs with total mass ranging from $10^4\, M_\odot$ to $10^7\, M_\odot$, for an illustrative redshift of $z=1$.
From the right panel (a) of Fig.~\ref{fig:snroft} we see that the smaller the total mass of the MBHB system, the higher the SNR of the early inspiral stage.
MBHBs with $M=10^4\, M_\odot$ and $M=10^5\, M_\odot$ at z=1 reach an SNR of $O(1)$ several years before merger, but then accumulate additional SNR more slowly than more massive sources in the last several months before merger.
However, the $M=10^4\, M_\odot$ MBHB does not merge within the frequency band of LISA, and thus its total SNR at merger is significantly lower than that of heavier systems, as shown in the left panel (b) of Fig.~\ref{fig:snroft}.
On the other hand, the $M=10^7\, M_\odot$ MBHB accumulates a negligible SNR during the early inspiral stage.
Until less than half a year before merger, its SNR is still below 1.
However, as shown in the left panel (b) of Fig.~\ref{fig:snroft}, in the last few months of observation, the SNR dramatically increases and becomes the loudest at merger if compared with smaller-mass systems. Note that all SNRs would be rescaled when considering different redshift values, and $z=1$ is rather optimistic for the loudness of most LISA sources. 

\begin{figure}[!htb]
    \centering
    \includegraphics[width=0.5\textwidth]{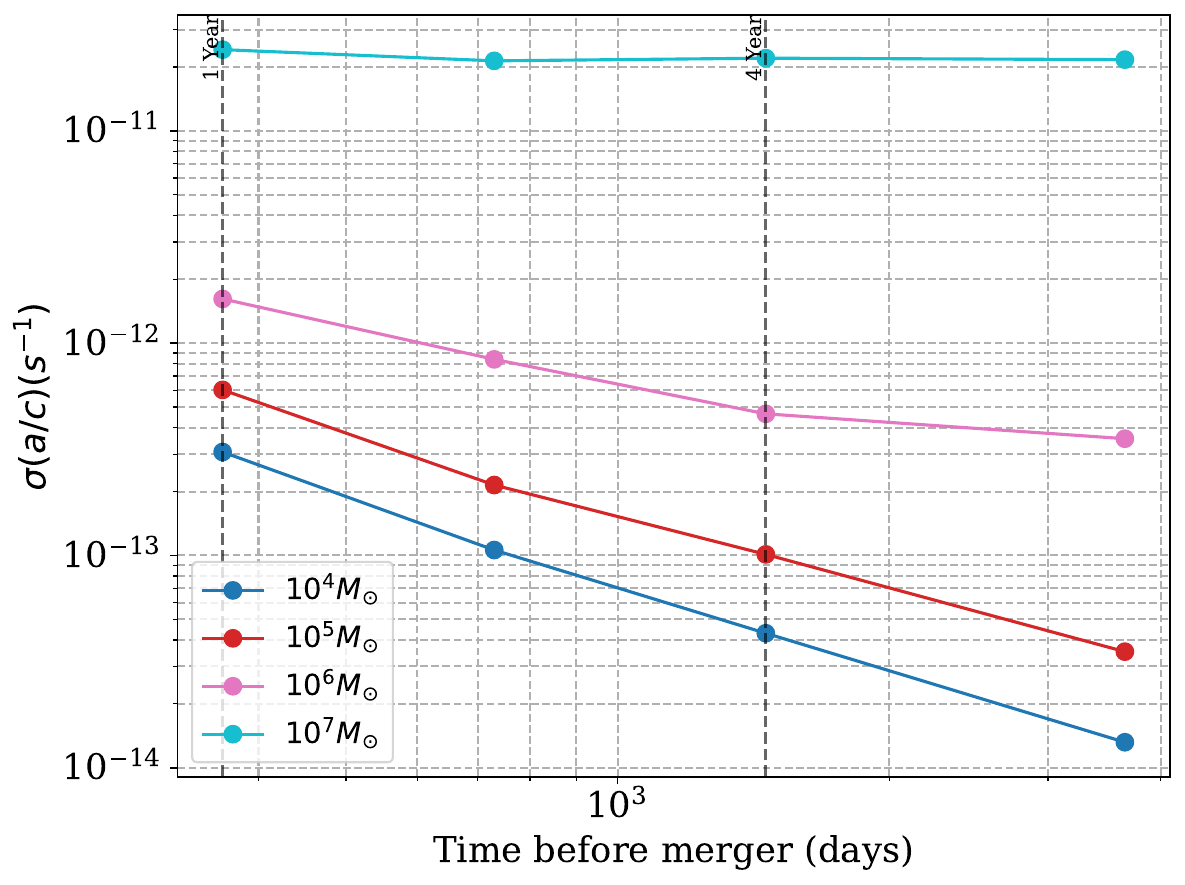} 
    \caption{Constraints at the 1-sigma level on the LOSA normalized by the speed of light, \(\sigma(a/c)\), as a function of the time before merger (in days). Each curve represents a MBHB system with a fixed set of intrinsic parameters (see details in the Table. \ref{other paras} in App.~\ref{appendix paras}) except for the total mass, which varies between curves as indicated in the legend. For all systems, the only varying parameter within each curve is the start time of the observation.
    }
    \label{fig:acc_single}
\end{figure}

The accumulation of SNR over time directly impacts how the LOSA is measured across MBHB signals.
At early times, the inspiral part of the signal might represent a cumulative SNR that remains below the detection threshold (say $\mathrm{SNR}=10$), while still bringing a significant contribution to the likelihood penalty of ignoring LOSA, according to Eq.~\eqref{eq:snr_likelihood_density}.
To show how the recovered uncertainty on the LOSA changes according to the observational time before merger and the total mass of the MBHB system, we compare four masses over signals ranging from 1 year to 10 years.
Fig.~\ref{fig:acc_single} presents the standard deviation $\sigma(a/c)$ as a function of the observational time before merger for systems with different total masses.
Lower-mass systems ($\lesssim 10^5\, M_\odot$) show a steady decrease in the uncertainty on the LOSA as more time before merger is added to the analysis.
This is in line with the previous SNR considerations: low-mass MBHBs slowly inspiral from the lower end of the LISA frequency band until their merger, providing a large number of cycles at SNR$\gtrsim1$ and lower frequencies from which relevant information on the LOSA can be inferred given its $-4$PN nature.
In contrast, for the $10^7 M_\odot$ system, most of the SNR is accumulated during the final weeks before merger, as the early inspiral phase yields negligible signal power in the LISA band.
Consequently, for high-mass MBHBs, extending the observational window to earlier times before merger has little impact on improving the measurement of the LOSA.
As shown in Fig.~\ref{fig:acc_single}, the constraint retrieved on the LOSA is roughly constant over time for these systems.

\begin{figure}[!htb] 
    \centering
    \includegraphics[width=0.5\textwidth]{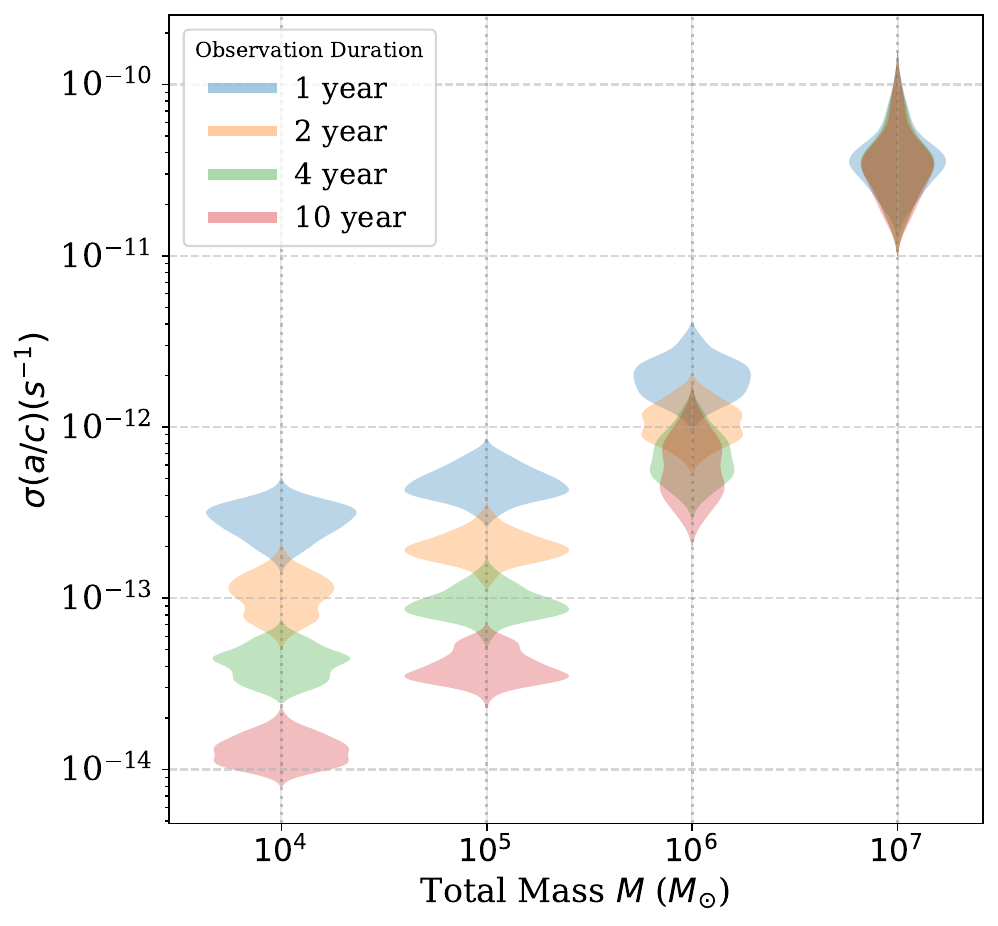} 
    \caption{The violin plot shows the relationship between the total mass $M$ (in solar masses $M_{\odot}$) of MBHBs and the standard deviation of LOSA $\sigma_{a/c}$, for four observation lengths (starting at 1, 2, 4, and 10 years before merger). Each violin plot is constructed from 100 random realizations for a given total mass, with fixed mass ratio $q=3$ and distance $z=1$, while other parameters (e.g., spins, orbital inclination, etc.) vary within the ranges shown in Table \ref{tab:random} of App.~\ref{appendix paras}.
    }
    \label{fig:acc_violin}
\end{figure}

To further investigate the influence of the total mass and observational time on the LOSA measurement, we generate an ensemble of 100 random realizations for each MBHB with fixed total binary mass, mass ratio ($q=3$) and distance ($z=1$), while randomly drawing other waveform parameters (spins, inclination, etc.) from physically motivated distributions.
These distributions are specified in Table \ref{tab:random} in App.~\ref{appendix paras}.
The results are reported as violin plots of $\sigma_{a/c}$ in Fig.~\ref{fig:acc_violin}, where the violin distributions show the variability across systems with different parameters.
The overall trend is consistent with the single-event results presented in Fig.~\ref{fig:acc_single}: $\sigma_{a/c}$ continuously decreases with increasing observation duration and exhibits a strong dependence on the total mass, with lower masses providing the best constraints; the higher the MBHB total mass, the less marked is the effect of the time of observation before merger.
For the heaviest systems we consider, with total mass $M=10^{7} M_{\odot}$, the observation duration  does not affect recovered uncertainties on the LOSA.
Fig.~\ref{fig:acc_violin} also shows that differences in the injected values of other waveform parameters, beyond the total mass and the time of observation, provide only moderate variations in the resulting LOSA uncertainty.

As an example of a small-SNR early inspiral contributing to the LOSA measurement, a close inspection of Figs.~\ref{fig:snroft} and~\ref{fig:acc_single} for the case $M_{\text{total}} = 10^6 \, M_{\odot}$ shows that the inspiral-only SNR is only about 2--3 between the 1-year and 2-year time marks, while the corresponding standard deviation $\sigma_{a/c}$ is reduced by approximately 50\% when changing the observation window.

\begin{figure*}[!htb] 
    \centering
    \includegraphics[width=\textwidth]{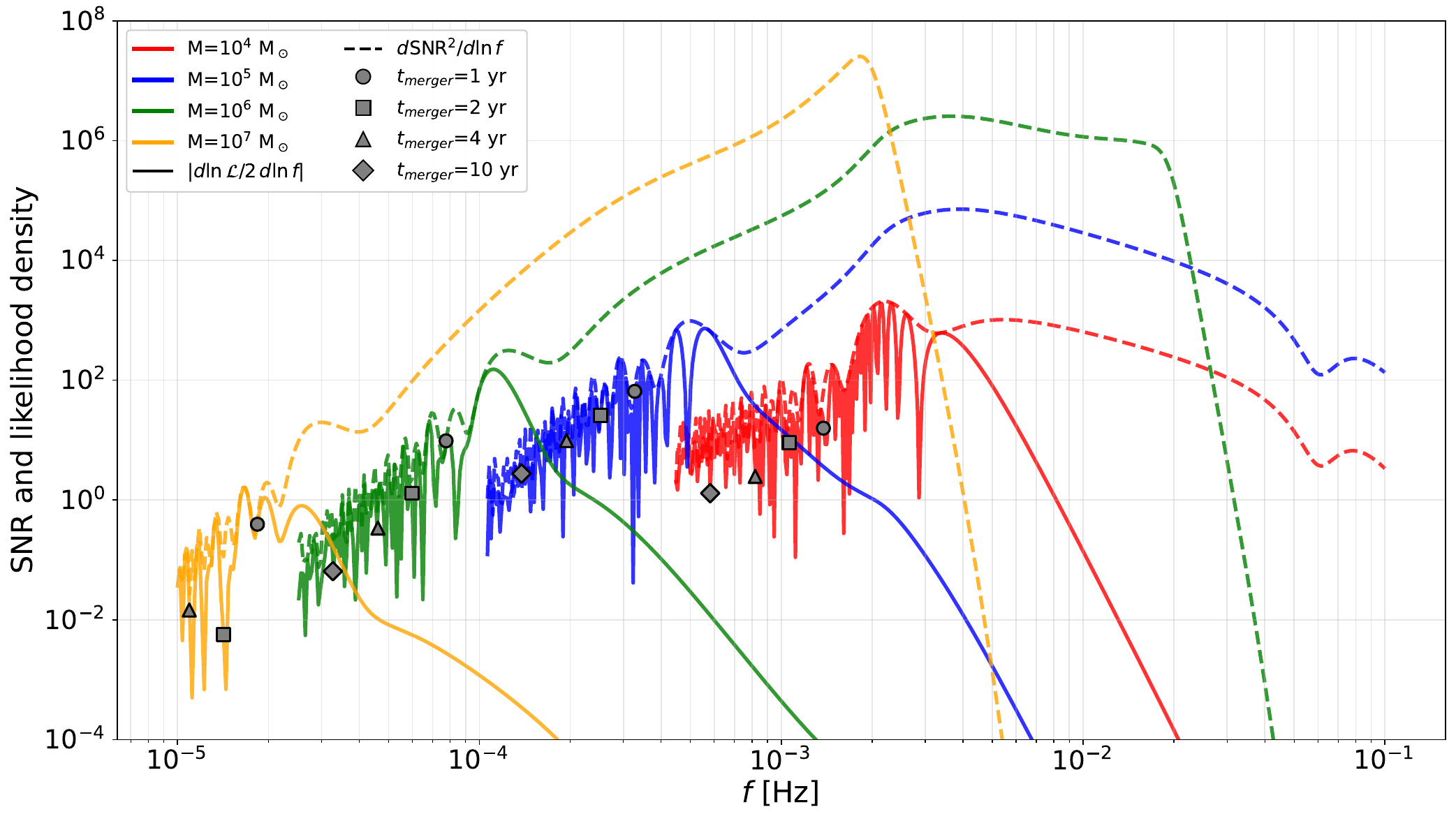} 
    \caption{Comparison of the frequency-resolved contributions to the logarithmic derivative of the likelihood function, $|d\ln\mathcal{L}/(2\,d \ln f)|$ (solid lines), and to the square SNR $ \rho^2$, $|d \rho^{2}/(d\ln f)|$ (dotted lines), as functions of gravitational wave frequency $f$, for massive black hole binaries with total masses $M = 10^4$, $10^5$, $10^6$, and $10^7\, M_\odot$ (red, blue, green, and orange, respectively). Symbols on the curves mark the frequencies corresponding to different time-to-merger values. In the likelihood calculation, the injection assumes a GR waveform with zero LOSA, while the template waveform is affected by a LOSA $a = 0.01\,\mathrm{m\,s}^{-2}$ ($a/c = 3.33\times 10^{-11}\,\mathrm{s}^{-1}$). We can visualize which part of the signal contributes most to the SNR and likelihood densities, with the LOSA-induced dephasing causing the bound Eq.~\eqref{eq:snr_likelihood_density} to become quickly saturated at low frequencies. 
    }
    \label{fig:likehoodandsnr}
\end{figure*}

Fig.~\ref{fig:likehoodandsnr} is designed to illustrate the contribution from different frequency bands to the log-likelihood function and the SNR. This figure gives the SNR density $d\rho^2/d\ln f$ and (half) the log-likelihood density $d(-\ln \mathcal{L})/(2d\ln f)$ as a function of frequency, following Eq.~\eqref{eq:snr_likelihood_density}.
In computing the log-likelihood, the injected waveform is taken to have zero LOSA, while the template is affected by a LOSA $a = 0.01\,\mathrm{m\,s}^{-2}$ corresponding to $a/c = 3.33\times 10^{-11}\,\mathrm{s}^{-1}$, and the four curves correspond to four example systems with different total masses.
Fig.~\ref{fig:likehoodandsnr} shows that the bound in Eq.~\eqref{eq:snr_likelihood_density} holds: the half-log-likelihood is bounded by the SNR density. It also confirms our intuition in showing explicitly in which part of the signal the information about the LOSA is located: the merger part of the signal, at high frequencies, brings little information on the LOSA; going from right to left in the figure, we reach a point where the LOSA-induced dephasing becomes significant, and we quickly transition to a regime where the bound for the log-likelihood density is saturated. This is yet another illustration of the importance of very early times, or low frequencies, for the LOSA measurement.

We stress that this point-wise computation of the likelihood density is a simplified view of parameter estimation, since we vary only the acceleration parameter, keeping others fixed, and ignoring possible correlation between parameters; however, Fig.~\ref{fig:likehoodandsnr} is already useful in understanding the difficulties of the Fisher matrix in properly simulating parameter estimation, which we will expose in App.~\ref{app:fisher_low_freq}. In short, the linearized-signal approximation at the heart of the Fisher matrix formalism does not apply to the rapidly increasing dephasing at low frequencies, leading to overshooting the bound of Eq.~\eqref{eq:snr_likelihood_density}, and causing a spurious overestimation of the precision achieved for the LOSA measurement.

\subsection{Effect of LOSA on measurement of other waveform parameters}
\label{subsec:otherparas}

\begin{figure*}[!htb] 
    \centering
    \includegraphics[width=\textwidth]{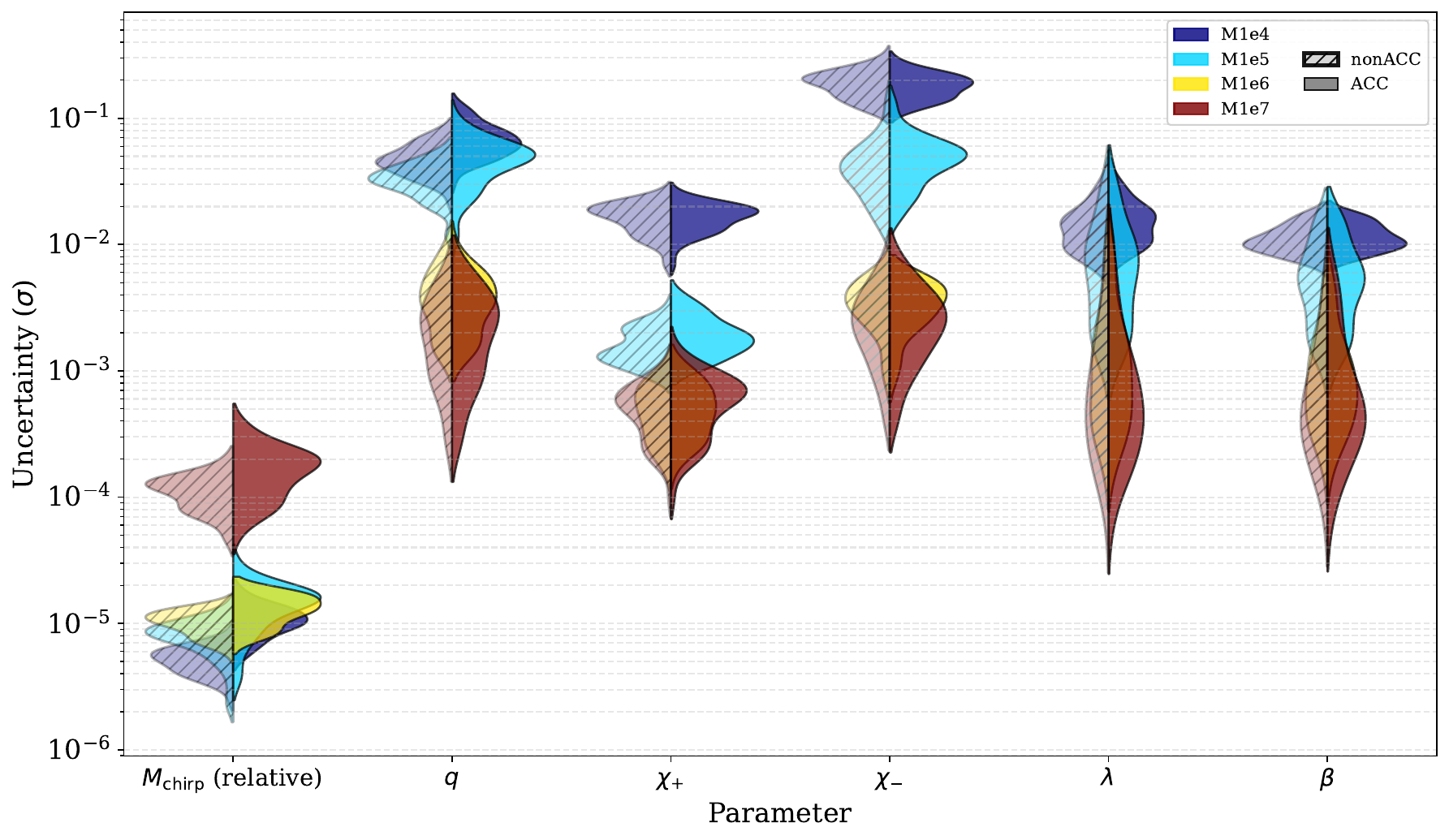} 
    \caption{
    Comparison of the recovered precision on waveform parameters by analysing a zero LOSA injection with a waveform containing LOSA terms (\textsc{ACC}) or not (\textsc{nonACC}).
    The violin plots show the uncertainty of the posterior distributions for six source parameters ($M_{\mathrm{chirp}}, q, \chi_{+}, \chi_{-}, \lambda, \beta$) on a logarithmic scale. For $M_{\mathrm{chirp}}$, the displayed uncertainty is the fractional standard deviation ($\sigma/M_{\mathrm{chirp}}$), while for all other parameters it is the absolute standard deviation ($\sigma$).
    Results are displayed for the two waveform models (ACC and nonACC) across different total masses, with each pair of violins split to compare the two models.
    Each half-violin represents the distribution derived from 100 random simulations with fixed total mass, mass ratio $q=3$, 4-year duration, and other parameters varying within plausible ranges shown in Table \ref{tab:random} in App.~\ref{appendix paras}. 
    }
    \label{fig:other paras}
\end{figure*}

In Fig.~\ref{fig:other paras} we compare the recovered precision on a subset of waveform parameters by injecting a zero LOSA signal and analysing it with two different templates: one including LOSA (ACC template) and the other excluding LOSA (nonACC template).
Although the LOSA parameter is set to zero (\(a/c=0\)) in the injected signals, the inclusion of the LOSA term in the recovery model still broadens, on average, the uncertainties on the chirp mass ($\mathcal{M}$) and the mass ratio ($q$), especially for low-total-mass systems.
Specifically, for the $10^4$ and $10^5\, M_\odot$ configurations, the \textsc{ACC} model (right half of the violins) yields noticeably larger uncertainties compared to the \textsc{nonACC} baseline (left half).
This degradation in precision is particularly visible for the chirp mass and mass ratio, intrinsic parameters that affect the overall phasing of the inspiral. This illustrates that the additional degree of freedom introduced by the LOSA correction degrades the inference of the intrinsic binary parameters, particularly for total masses of $10^4$ and $10^5\, M_\odot$.
Other parameters, however, are generally unaffected by the inclusion of the LOSA.

\subsection{Bias induced by the LOSA on other parameters}
\label{bias}

\begin{figure}[!htb] 
    \centering
    \includegraphics[width=0.5\textwidth]{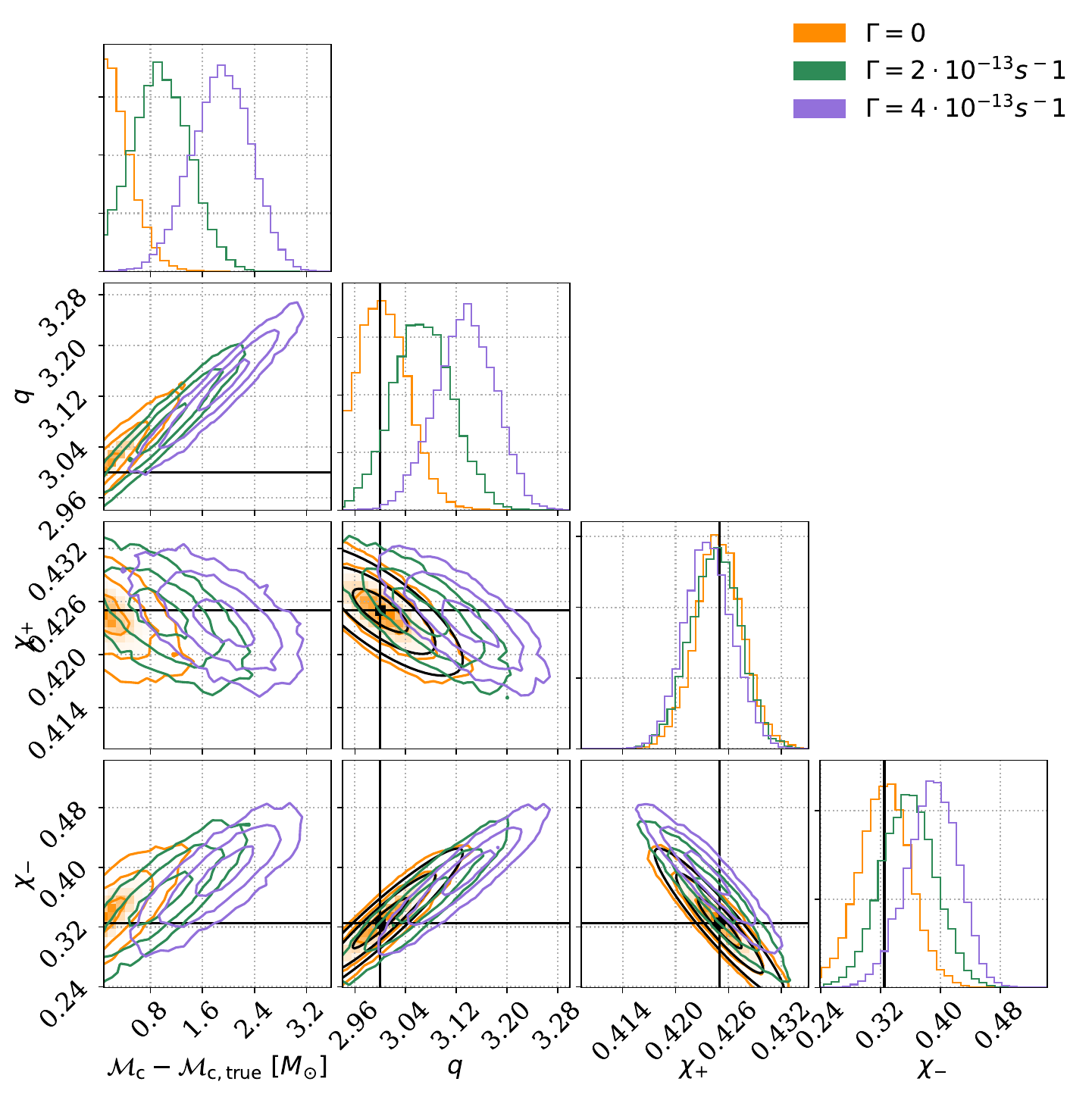} 
    \caption{Corner plot illustrating the parameter estimation bias induced by neglecting LOSA in parameter estimation analyses. The simulated source is a MBHB system with a total mass of \(10^5 M_{\odot}\) and a mass ratio of $q=3$ (see other parameters in Table.~\ref{other paras} in App.~\ref{appendix paras}), for an observation of 4 years.
    The corner plot shows the one-dimensional marginalized posteriors and two-dimensional confidence contours for the chirp mass \(\mathcal{M}_c\), mass ratio \(q\), and effective spins \(\chi_+\) and \(\chi_-\). The three colors correspond to different values of the LOSA parameter $\Gamma$ imposed on the waveform: orange (\(\Gamma=0\), a zero LOSA injection), green (\(\Gamma = 2 \times 10^{-13} \, \text{s}^{-1}\approx1\sigma\)), and purple (\(\Gamma = 4 \times 10^{-13} \, \text{s}^{-1}\)). The shift of the contours relative to the orange injection reveals the systematic bias introduced by the unmodeled LOSA effect.
    }
    \label{fig:time_corner}
\end{figure}

In this section, we investigate the bias induced on the waveform parameters when a LOSA is present, but the signal is analysed with a template ignoring LOSA.

The presence of an unmodeled LOSA fundamentally distorts the parameter estimation of MBHBs. To explicitly illustrate this systematic bias, we inject a 4-year-duration MBHB signal with total mass \(10^5\, M_{\odot}\) and mass ratio \(q=3\) (with the other parameters listed in Table~\ref{other paras} of App.~\ref{appendix paras}) that is influenced by the LOSA (\(a>0\)), and we recover it using a non-ACC template that assumes \(a=0\). The results are reported in Fig.~\ref{fig:time_corner}.

As Fig.~\ref{fig:time_corner} shows, the severity of the biases found in such an experiment will be dependent on the strength of the LOSA that we inject; here we choose values that lie around the detection threshold for the LOSA effect: green corner (\(\Gamma = 2 \times 10^{-13} \, \text{s}^{-1}\approx1\sigma\)), and purple corner (\(\Gamma = 4 \times 10^{-13} \, \text{s}^{-1}\)).

The most severe bias is observed in the chirp mass \(\mathcal{M}_c\). Compared to the true zero LOSA injection (orange contours), the recovered posterior distributions shift prominently toward higher values as the LOSA increases to higher values.
Crucially, this systematic bias is not confined to the chirp mass alone; the mass ratio \(q\) exhibits a concurrent upward shift, indicating a strong degeneracy between the redshifted phase evolution induced by LOSA and the intrinsic mass parameters of the binary.

In addition to the mass parameters, the LOSA also biases the inference of the spin configuration. While in this example the effective spin \(\chi_+\) remains relatively stable in its central value, the two-dimensional posterior contours reveal a bias in the anti-correlation between \(\chi_+\) and \(\chi_-\) as \(a\) increases, corresponding to a small bias on \(\chi_-\).
Overall, when LOSA is neglected, some of the recovered waveform parameters deviate from their true values, posing a significant risk of misinterpretation for sources in dense environments.

\subsection{Validity of the constant LOSA approximation}
\label{subsec:validity_constant}

\begin{figure}[!htb] 
    \centering
    \includegraphics[width=0.5\textwidth]{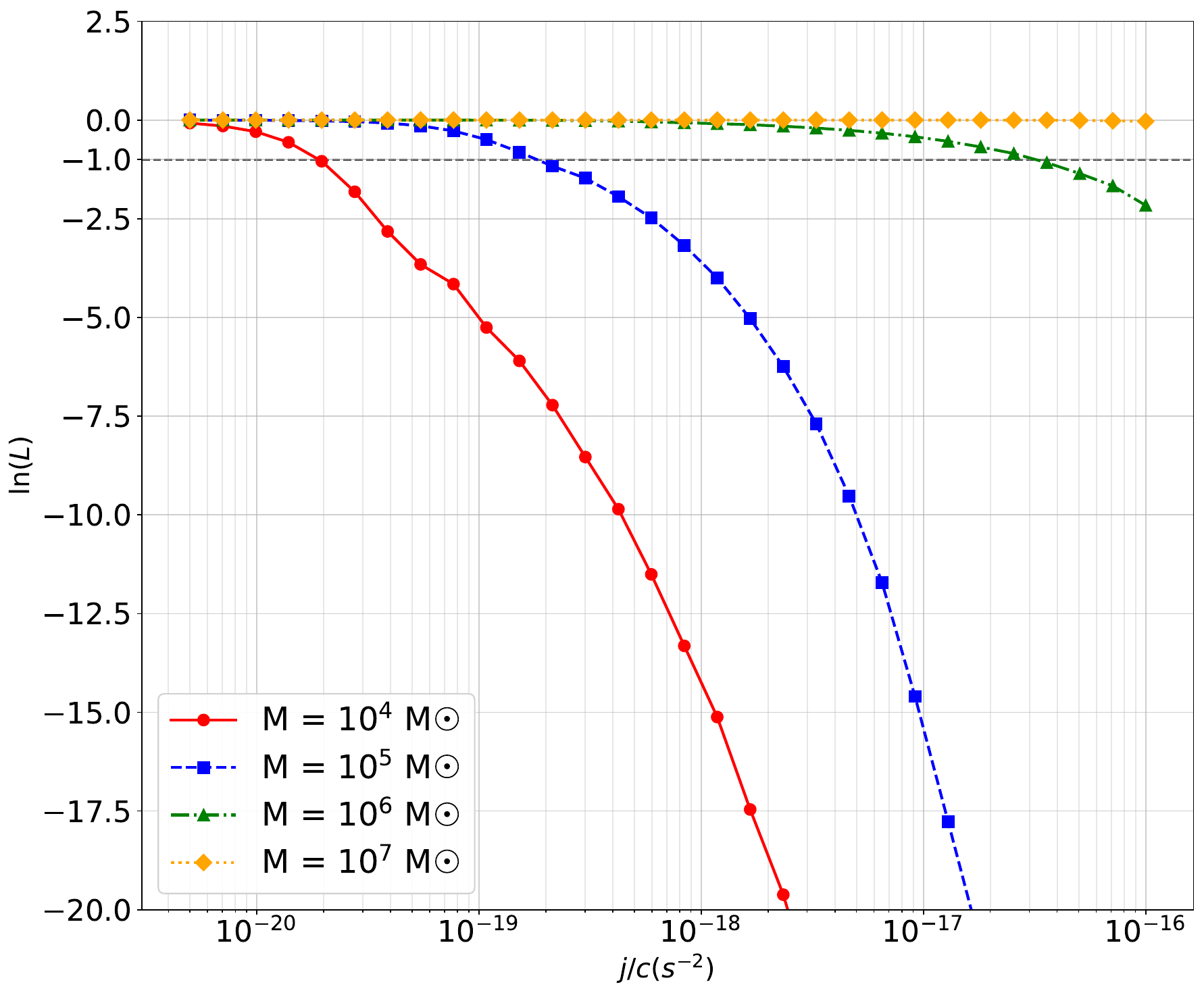} 
    \caption{Log-likelihood of template matching as a function of jerk for MBHB systems of different masses. The horizontal axis shows the jerk (normalized to the speed of light $c$, in $s^{-2}$), and the vertical axis shows $\log\mathcal{L}$ with an injection including the jerk and a template that sets it to zero (negative values, with values closer to zero indicating a better template match). The four curves correspond to MBHB systems with total masses of $10^7\, M_\odot$ (orange), $10^6\, M_\odot$ (green), $10^5\, M_\odot$ (blue), and $10^4\, M_\odot$ (red); the other parameters are reported in Table \ref{other paras} in App.~\ref{appendix paras}.
    All templates used to compute the log-likelihood assume constant LOSA $a/c = 10^{-12}\,\mathrm{s}^{-1}$ (zero jerk), while different jerk values are injected in the signal as reported on the $x$-axis. 
    }
    \label{fig: dadt}
\end{figure}

In this section, we perform simple checks to understand the regime of validity of the constant LOSA approximation employed throughout our study.

To evaluate the impact of considering the LOSA in the gravitational wave template as constant, we compute the log-likelihood (\(\log\mathcal{L}\)) obtained by injecting a reference signal with a non-constant acceleration in the form of a non-zero jerk, and recovering it with a template that ignores it.
In Fig.~\ref{fig: dadt}, we present the result across different total MBHB masses \(M\) from \(10^4\,M_\odot\) to \(10^7\,M_\odot\). The injection and template are both constructed with a non-zero LOSA $a/c = 10^{-12}\,\mathrm{s}^{-1}$. The templates have $j=0$ (zero jerk), while the injected signals incorporate a LOSA variation at the next-to-leading order (non-zero jerk), shown on the $x$-axis. Log-likelihood values closer to zero indicate a better match.
In other words, the closer the log-likelihood is to zero, the more reliable a signal is when ignoring the jerk; log-likelihood values away from zero indicate that jerk corrections cannot be ignored and potentially that $j$ can be measured. This numerical experiment essentially amounts to a 1D parameter estimation, and ignores correlations with other parameters that are here kept fixed.

The results reported in Fig.~\ref{fig: dadt}
reveal a strongly mass-dependent behavior: for the most massive system (\(M = 10^7\, M_\odot\), orange curve), the log-likelihood remains close to zero across the entire tested jerk range, indicating that the constant LOSA template provides an excellent fit even when the jerk is ignored.
Conversely, for the less massive system, the sensitivity to the jerk corrections cannot be ignored.
For the intermediate mass systems (\(10^6\,M_\odot\) in green and \(10^5\,M_\odot\) in blue), \(\log\mathcal{L}\) begins to drop as the jerk magnitude increases. This trend is most pronounced for the lowest-mass system (\(M = 10^4\, M_\odot\), red curve), where the log-likelihood exhibits a decline beyond \(j/c \approx 10^{-19}\,\text{s}^{-2}\).
We set the threshold to determine the applicable range of the constant acceleration assumption for $\log\mathcal{L}= -1$.
All jerk values below this threshold are considered negligible.
In Sec.~\ref{MCMC}, such a threshold will be expressed as a function of the third body's mass and its separation from the MBHB, indicating for which values of these quantities our constant LOSA approximation applies.

\section{Astrophysical implications}
\label{MCMC}

\begin{figure*}[!htb] 
    \centering
    \includegraphics[width=0.8\textwidth]{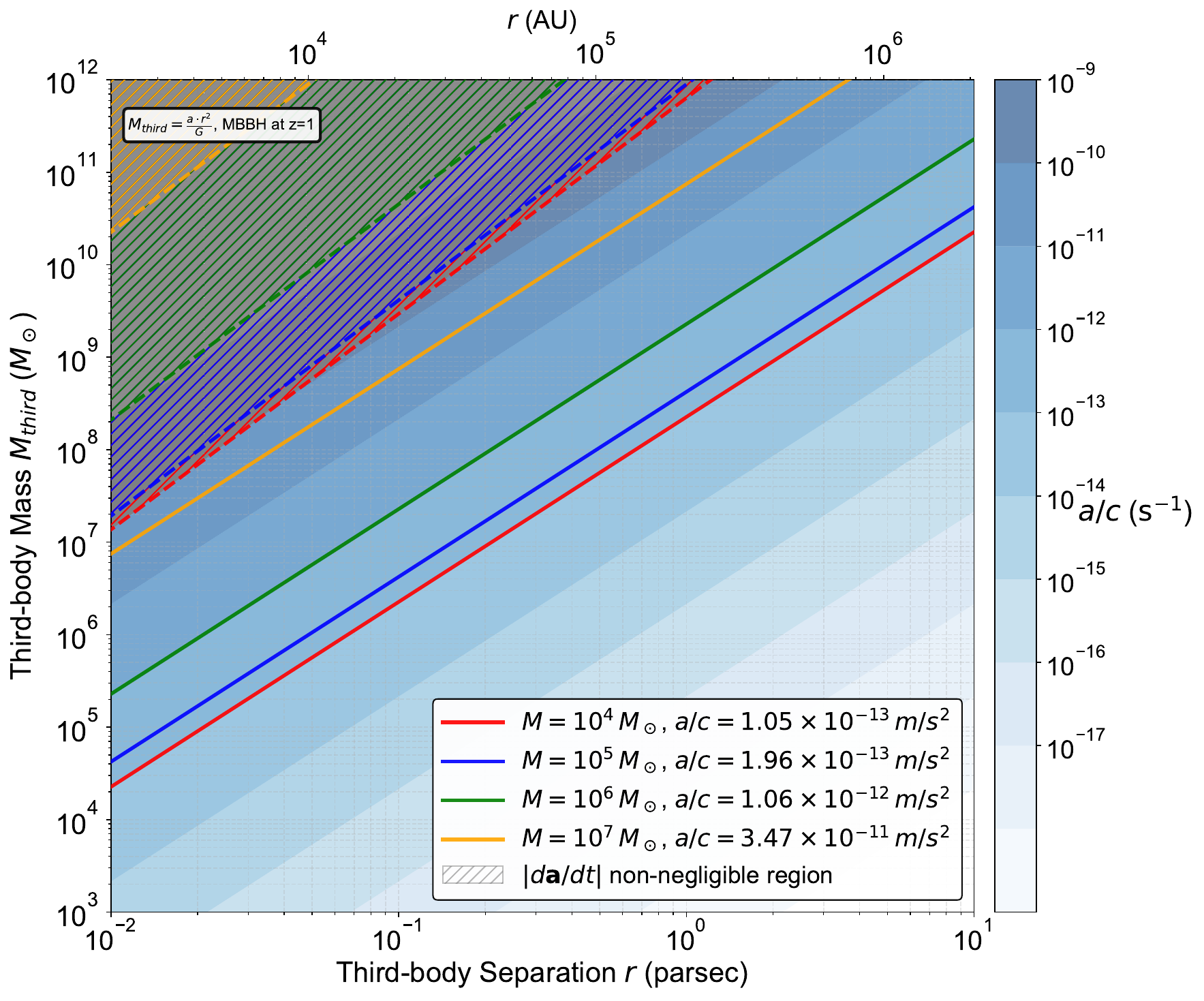} 
    \caption{Detectability threshold with LISA for a triple system composed of an inner MBHB at redshift \(z=1\) and a third MBH. The plot shows the required mass \(M_{\text{third}}\) of the third body (vertical axis) and its separation \(r\) from the MBHB's center of mass (horizontal axis) to induce a detectable LOSA, assuming perfect alignment with the line-of-sight (maximum phase of the acceleration induced by the outer orbit). The colored solid lines mark the detection lower limit for  4 years of observation for 4 typical MBBH total masses: red (\(10^4\, M_\odot\)), blue (\(10^5\, M_\odot\)), green (\(10^6\, M_\odot\)), and orange (\(10^7\, M_\odot\)). A third body located above a specific colored line produces a detectable LOSA for the corresponding MBHB mass at \(z=1\). The background shades of blue indicate levels of constant acceleration for reference. The hatched region in the upper left represents the parameter space where the induced acceleration varies significantly over the observation time (the \(|da/dt|\) non-negligible region), meaning the acceleration cannot be treated as constant; see Sec.~\ref{subsec:validity_constant}.   
    }    
\label{fig:third bodyr}
\end{figure*}

In this section we discuss the astrophysical implications of the results reported above.
In particular, we translate the values of the LOSA that LISA will be able to measure to separations and masses of the third MBH orbiting the MBHB source as follows:

\begin{equation}
    a_\parallel=(GM_{\rm third}/r^2)\,\hat{l}\cdot\hat{r},
\label{amr}
\end{equation}
where $M_{third}$ is the tertiary mass, r is the separation between the CoM of the binary and the third body, $\hat{l}$ is a unit vector along the line of sight, and $\hat{r}$ is a unit vector pointing from the binary CoM to the tertiary. 

The results reported in Fig.~\ref{fig:third bodyr} are derived from the posterior standard deviations $\sigma(a/c)$ of the LOSA parameter shown in Fig.~\ref{fig:acc_violin}, which are taken as a proxy for the minimum detectable LOSA thresholds for LISA. 
According to Eq.~\eqref{amr}, a given LOSA value corresponds to a straight line in the logarithmic parameter space of third-body mass $M_{\rm third}$ versus separation $r$. 
Since Fig.~\ref{fig:acc_violin} shows that MBHB systems with different total masses have different minimum detectable LOSA limits, we mark these corresponding thresholds with solid lines of different colors in Fig.~\ref{fig:third bodyr}. 
The blue-shaded background indicates the gradient direction of the LOSA. 
For a MBHB of a given total mass, any third body located above the corresponding colored line produces a detectable imprint on the GW signal via the induced LOSA.

For example, for a MBHB at $z=1$ with total mass \(10^4 M_\odot\) or \(10^5 M_\odot\) (red and blue lines), the method can constrain the LOSA induced by a third \(\sim 10^8 M_\odot\) MBH out to separations \(\sim 1\,\text{pc}\).
MBHBs with masses at the limits between massive and intermediate-mass black holes (IMBHs) ($\sim 10^4\, M_\odot$) could thus represent an ideal target to detect the LOSA with LISA.
IMBHs are believed to originate in dense environments such as globular clusters or AGNs (see e.g.~\cite{McKernan:2012rf,McKernan:2014oxa,Rasskazov:2019tgb,2024Sci...384.1488F}), which usually harbour a central MBH.
Although the rates of IMBHs are unknown at present, if the production of IMBHs toward the higher end of their mass range is enhanced in such dense environments for any reason, then we may expect LISA to be able to detect the LOSA induced by any sufficiently massive MBH which lies close enough to the IMBH binary (see also \cite{Inayoshi:2017hgw,Randall:2018lnh,Wong:2019hsq}).

On the other hand, from Fig.~\ref{fig:third bodyr} we infer that for MBHBs with total mass \(10^6\, M_\odot\) (\(10^7\, M_\odot\)) at $z=1$, only a third body with a mass of at least \(10^9\, M_\odot\) (\(10^{10}\, M_\odot\)) at separations (\(\sim 1\,\text{pc}\)), can cause a large enough LOSA to be detectable.

This means that MBHBs at the higher mass end of the LISA detectability range can show signs of the LOSA only if a supermassive black hole (\(M \gtrsim 10^9 M_\odot\)) lurks nearby.

Supermassive black holes are believed to be abundant in the universe as they are commonly found in the center of massive galaxies \cite{Richstone:1998ky, Kormendy:2013dxa}. It is well known that at separations of \(\sim 1\,\text{pc}\), the standard astrophysical mechanisms supposed to harden the MBHB cease to be effective, possibly leading to several MBHs stalling at such distances (the so-called final parsec problem).

Hierarchical triplets around massive black hole binaries (MBHBs) can drive coalescence through two distinct but complementary channels: secular Kozai--Lidov (K--L) oscillations and chaotic three-body interactions. In the K--L channel, the third body remains on a wide, inclined outer orbit while the inner binary undergoes periodic eccentricity excitation; in the chaotic channel, the hierarchy breaks down, and the system enters a phase of strong resonant encounters. Both channels are regulated by the ratio between the outer pericentre distance and the inner semi-major axis, and both require the system to be sufficiently compact that the outer perturber can exchange energy and angular momentum with the inner binary. A large suite of numerical integrations has quantified the stability boundary of hierarchical triplets and provided the criterion that separates stable hierarchical configurations from unstable ones \cite{2018triplestability}. MBH triplets followed through 2.5 PN order in spherical stellar potentials have shown that a significant fraction of otherwise stalled MBHBs, of order 20--30 per cent, are driven to coalescence within a Hubble time, with merger times peaking around 300 Myr and residual eccentricities near plunge of order 0.1 \cite{Bonetti:2017dan,li2026RAMCOAL}. Together, these results establish a coherent picture in which K--L and chaotic triplet channels both operate within the same hierarchy, with the stability criterion marking the transition between the secular and chaotic regimes \cite{2018triplestability}.

To demonstrate the possibility that LISA may be able to detect MBHBs with LOSA, consider a triplet with an outer mass of $10^{7}\, M_{\odot}$, an inner binary total mass of $10^{6}\, M_{\odot}$ (equal-mass), an inner semi-major axis of $0.010$ pc, and an outer eccentricity of $0.05$, adopting the standard assumptions of a circular inner binary and a coplanar prograde geometry. Applying the stability criterion with the coefficient recommended for $N$-body experiments, the resulting stability threshold corresponds to an outer pericentre of approximately $0.040$ pc (see calculation details in \cite{2018triplestability}). In other words, for this configuration the hierarchy remains stable as long as the third body stays outside roughly $0.040$ pc, and becomes dynamically unstable once the outer orbit shrinks below this value.

The physical significance of this threshold is that it separates two qualitatively different evolutionary paths. If the MBHB enters the gravitational-wave (GW) dominated regime while the third body is still on an orbit wider than the threshold, the triplet remains hierarchical. If, instead, the system evolves below the threshold, the hierarchy becomes chaotic, and after the ensuing three-body interaction the MBHB may enter the GW-dominated phase with the third body on a much tighter orbit, for example at about $0.01$ pc ($\Gamma \approx 4.63 \times 10^{-11} \text{s}^{-1}$), comparable to the inner semi-major axis itself. 
The key point is that the two paths predict different distributions of the outer semi-major axis at the epoch when the inner binary enters the LISA band: the hierarchical path leaves the third body at roughly $0.040$ pc or more, which means $\Gamma \leq 2.90 \times 10^{-12}\text{s}^{-1}$ (for this system, the low limit of $\Gamma$ is $1.06 \times 10^{-12}\text{s}^{-1}$ ), so we may observe LOSA in both cases.

We note that studies have pointed out that a considerable fraction of MBHB mergers in the LISA frequency band may be actually induced by the presence of a third MBH \cite{Bonetti:2016eif,Bonetti:2017dan,Bonetti:2017lnj,Bonetti:2018tpf}.
When coupled to a semi-analytic galaxy formation model \citep{Bonetti:2018tpf}, this triplet channel contributes $\sim 3$--$40\%$ of MBHB mergers under standard assumptions, rising to 100\% in stalling scenarios and corresponding to 10--70 LISA MBHB merger detections per year.

Another piece of information reported in Fig.~\ref{fig:time_corner} is the validity region of our constant LOSA approximation in the third-body mass-separation parameter space.
To give an example, the outer orbital period of a $M=10^5\, M_\odot$ MBHB orbiting a $m_3 = 10^8\, M_\odot$ MBH at a separation of $a_{\rm out} = 1\,$pc is
\begin{equation}
    P_{\rm out} \sim 2\pi\sqrt{\frac{a_{\rm out}^3}{Gm_3}} \approx 10^4\,\text{yr} \,,
\end{equation}
which is much larger than the LISA mission lifetime ($\sim 10\,$yr at most).
Consequently, during any realistic observational window, the MBHB effectively sits at a fixed position in the third MBH potential, and the binary inner system does not exhibit measurable orbital-phase modulation due to the outer motion. 
Third body at closer separations and higher masses, however, may have much shorter orbits for which our constant LOSA approximation does not apply (see Sec.~\ref{subsec:validity_constant}).

The hatched region in the upper right corner of Fig.~\ref{fig:third bodyr} marks the parameter space where the acceleration rate of change \(|da/dt|\) is non-negligible. This region is defined as the set of parameters for which the log-likelihood satisfies \(\log\mathcal{L} < -1\) when comparing an injection and a template with and without jerk (see Sec.~\ref{subsec:validity_constant}).
Within this region, the induced acceleration varies significantly over the observational timescale, meaning the constant acceleration assumption employed in our analysis is no longer valid.
So the third body in this area can still be detected, but the constant LOSA waveform used in our study does not constitute a suitable representation of the GW signal.
More complex waveform models must be employed in such cases, similar to the ones already applied to study less massive binary systems \cite{Sberna:2022qbn,2021PhRvL.126j1105T,Katz:2022izt}.

\section{Conclusion}
\label{sec:conclusions}

In this work, we have investigated the detectability of the LOSA in GW signals from MBHBs observable by LISA, examining its impact on parameter estimation and its utility for constraining nearby third bodies.

Our analysis reveals a pronounced dependence on both the SNR evolution and the total MBHB mass.
As illustrated in Fig.~\ref{fig:snroft}, lower-mass systems (total mass in the range $10^4$--$10^5\, M_\odot$) accumulate SNR gradually over extended early inspiral phases. Because the LOSA-induced phase correction arises dominantly at the $-4\,\text{PN}$ order during the low-frequency inspiral, the precision of LOSA measurement improves substantially with longer observational times, and consequently for lower-mass systems (Figs.~\ref{fig:acc_single} and \ref{fig:acc_violin}).
For instance, for a $10^6\, M_\odot$ MBHB, extending the observation window from one to two years reduces the uncertainty on the LOSA by approximately $50\%$, despite only a modest SNR being contained in that early inspiral. 

This improvement underscores the critical importance of the low-frequency portion of the LISA band. The LOSA-induced phase shift accumulates most significantly at very early time, and in the frequency range between $10^{-5}\,\text{Hz}$ and $10^{-4}\,\text{Hz}$, where the $-4\,\text{PN}$ correction to the GW phase is most pronounced.
Data in this range of frequencies contain the bulk of the LOSA information for the most favorable, low-mass systems and are therefore essential for the measurement of this effect. 
Even though individual frequency bins in this regime may carry low SNR, their coherent integration over multi-year baselines provides the majority of constraining power on the LOSA.
Consequently, the duration of the LISA mission (nominally of roughly 4 years, with a possible extension to 10 years) as well as the effective low-frequency cutoff of LISA---whether determined by instrumental sensitivity limits or by data analysis choices--- will both directly impact the achievable precision of LOSA measurements.

We also compared parameter estimation performance between templates containing and not containing the LOSA deviations.
When signals with $a=0$ are analyzed with a model including the LOSA terms, the additional degree of freedom broadens the posterior distributions for the chirp mass $\mathcal{M}_c$ and the mass ratio $q$, particularly for $10^4$ and $10^5\, M_\odot$ total mass MBHB systems (Fig.~\ref{fig:other paras}).
On the other hand, neglecting a non-zero LOSA in the recovery waveform template leads to significant systematic shifts in $\mathcal{M}_c$ and $q$ if the strength of the LOSA is near the detectability threshold (Fig.~\ref{fig:time_corner}).

The assumption of constant acceleration was validated by injecting signals with non-zero jerk. As shown in Fig.~\ref{fig: dadt}, high-mass systems ($10^7\, M_\odot$) are largely insensitive to jerk corrections, while lower-mass systems ($10^4\, M_\odot$) exhibit degraded template matching when $\dot{a}/c \gtrsim 10^{-19}\,\text{s}^{-2}$. We required a threshold of $\log\mathcal{L}=-1$ to identify the regime where the constant-acceleration approximation remains sufficiently accurate.

Finally, we mapped the detectability of the LOSA to the third-body mass--separation plane.
For $10^4\,M_\odot$ and $10^5\,M_\odot$ MBHBs at $z=1$, LISA can constrain the LOSA of a $10^8\,M_\odot$ perturber out to $\sim 1\,\text{pc}$.
At the same separation scales, more massive MBHBs ($\gtrsim 10^6\, M_\odot$) require a more massive perturber ($\gtrsim 10^9\, M_\odot$) to create a detectable LOSA.
MBH companions at parsec scales from the MBHB could be common in dense environments for light systems ($\lesssim 10^5\, M_\odot$), and due to the final parsec problem in heavier ones ($\gtrsim 10^6\, M_\odot$).
Future investigations will be needed to predict the number of LOSA detections for LISA MBHBs, as well as to expand our analysis to include more complex waveform models, including, in particular, the MBHB eccentricity (see \cite{Pompili:2026tdf, Roy:2026duh, Roy:2026mco, Hendriks:2024zbu, Samsing:2024syt} for similar research with current ground-based GW observations).

In summary, the LOSA introduces a characteristic low-frequency phase signature that is most significant for low-to-intermediate-mass binary black holes. Properly modeling this effect is crucial both to avoid systematic biases in parameter estimation and to achieve a novel probe of massive perturbers in the environments of MBHB sources.

\begin{acknowledgments}

The authors are grateful to John Baker and Stanislav Babak for insightful and stimulating discussions that helped sharpen the arguments presented in this paper. Thanks are also due to Vasco Gennari and Mathieu Dubois for their generous technical assistance,  which substantially facilitated the completion of this work.
The authors acknowledge support from the French space agency CNES in the framework of LISA.

\textit{Software}:
We acknowledge usage of the following additional
\textsc{Python} 
packages for modeling, analysis, post-processing, and production of results throughout:
\textsc{matplotlib}~\cite{Hunter:2007},
\textsc{numpy}~\cite{Harris:2020},
\textsc{scipy}~\cite{Virtanen:2020}
.
\end{acknowledgments}

\appendix\label{appendix}

\section{Additional information on the simulations presented in this work}
\label{appendix paras}

\begin{table}[htbp]
\centering
\caption{Additional waveform injection parameters for the simulations underlying the results presented in Figs.\ref{fig:snroft}, \ref{fig:acc_single}, \ref{fig:likehoodandsnr}, \ref{fig:time_corner}, and \ref{fig: dadt}. All angles and the coalescence time shift refer to the SSB frame.}
\label{other paras}
\begin{tabular}{l l}
\hline\hline
Parameter & Value \\
\hline
$q$ & $3.0$ \\
$\chi_1$ & $0.5$ \\
$\chi_2$ & $0.2$ \\
$\iota$ & $\pi/3$ \\
$\phi$ & $1.2$ \\
$\lambda$ & $0.8$ \\
$\beta$ & $0.3$ \\
$\psi$ & $1.7$ \\
\hline\hline
\end{tabular}
\end{table}

Here we provide additional information on the simulations used to produce the main results of our study.
In Table \ref{other paras} we provide the list of injected parameters for the simulations underlying the results presented in Figs.~\ref{fig:snroft}, \ref{fig:acc_single}, \ref{fig:likehoodandsnr}, \ref{fig:time_corner}, and \ref{fig: dadt}.
\begin{table}[htbp]
\centering
\caption{The distributions of  source parameters for generating random binary in Figs.~\ref{fig:acc_violin} and \ref{fig:other paras}.}
\label{tab:random}
\begin{tabular}{@{}llcc@{}}
\toprule
\textbf{Parameter} & \textbf{Sampling way} & \textbf{Range} \\
$\chi_{1}$ & Uniform & $[-0.98,\; 0.98]$ \\
$\chi_{2}$ & Uniform & $[-0.98,\; 0.98]$ \\
 $\iota$ & $\propto \sin\iota$ & $[0,\; \pi]$ \\
 $\phi$ & Uniform & $[-\pi,\; \pi]$ \\
 $\lambda$ & Uniform & $[-\pi,\; \pi]$ \\
 $\beta$ & $\propto \cos\beta$ & $[-\pi/2,\; \pi/2]$ \\
 $\psi$ & Uniform & $[0,\; \pi]$ \\
\end{tabular}
\end{table}
In Table \ref{tab:random}, we present the physically motivated injection distributions to generate a representative population of synthetic signals for Figs.~\ref{fig:acc_violin}, \ref{fig:other paras}.
\begin{table}[htbp]
\centering
\caption{Prior distributions for MCMC parameters estimations in Figs.\ref{fig:snroft}, \ref{fig:acc_single}, \ref{fig:likehoodandsnr}, \ref{fig:time_corner}, and \ref{fig: dadt}.}
\label{tab:priors}
\begin{tabular}{@{}llcc@{}}
\toprule
\textbf{Parameter} & \textbf{Prior} & \textbf{Range} \\
$\mathcal{M}_c$  ($\mathrm{M}_\odot$)& Uniform & $[10^3,\; 10^7]$ \\
$q$ & Uniform & $[2.7,\; 3.3]$ \\
$\chi_{\rm p}$ & Uniform & $[-1,\; 1]$ \\
 $\chi_{\rm m}$ & Uniform & $[-1,\; 1]$ \\
 $\Delta t$ & Uniform & $[-600,\; 600]$ \\
$d_L$ & Uniform & $[10^3,\; 10^4]$ \\
 $\iota$ & $\propto \sin\iota$ & $[0,\; \pi]$ \\
 $\phi$ & Uniform & $[0,\; 2\pi]$ \\
 $\lambda$ & Uniform & $[0,\; 2\pi]$ \\
 $\beta$ & $\propto \cos\beta$ & $[-\pi/2,\; \pi/2]$ \\
 $\psi$ & Uniform & $[0,\; \pi]$ \\
 $a/c$ & Uniform & $[-10^{-8}, \; 10^{-8}]$ \\
\end{tabular}
\end{table}
In Table \ref{tab:priors} we expose all the prior distributions used to perform Bayesian parameter estimation in the simulations underlying Figs.~\ref{fig:snroft}, \ref{fig:acc_single}, \ref{fig:acc_violin}, \ref{fig:likehoodandsnr}, \ref{fig:time_corner}, \ref{fig:other paras} and \ref{fig: dadt}. In practice, the chirp-mass prior is further tailored to each fixed mass, spanning one order of magnitude above and below the true total mass. The initialization of the MCMC chains is not done by drawing from the priors but by drawing from the Fisher matrix prediction for the posterior covariance.

\section{full corner plot}
\label{full corner}

\begin{figure*}
    \centering
    \includegraphics[width=\textwidth]{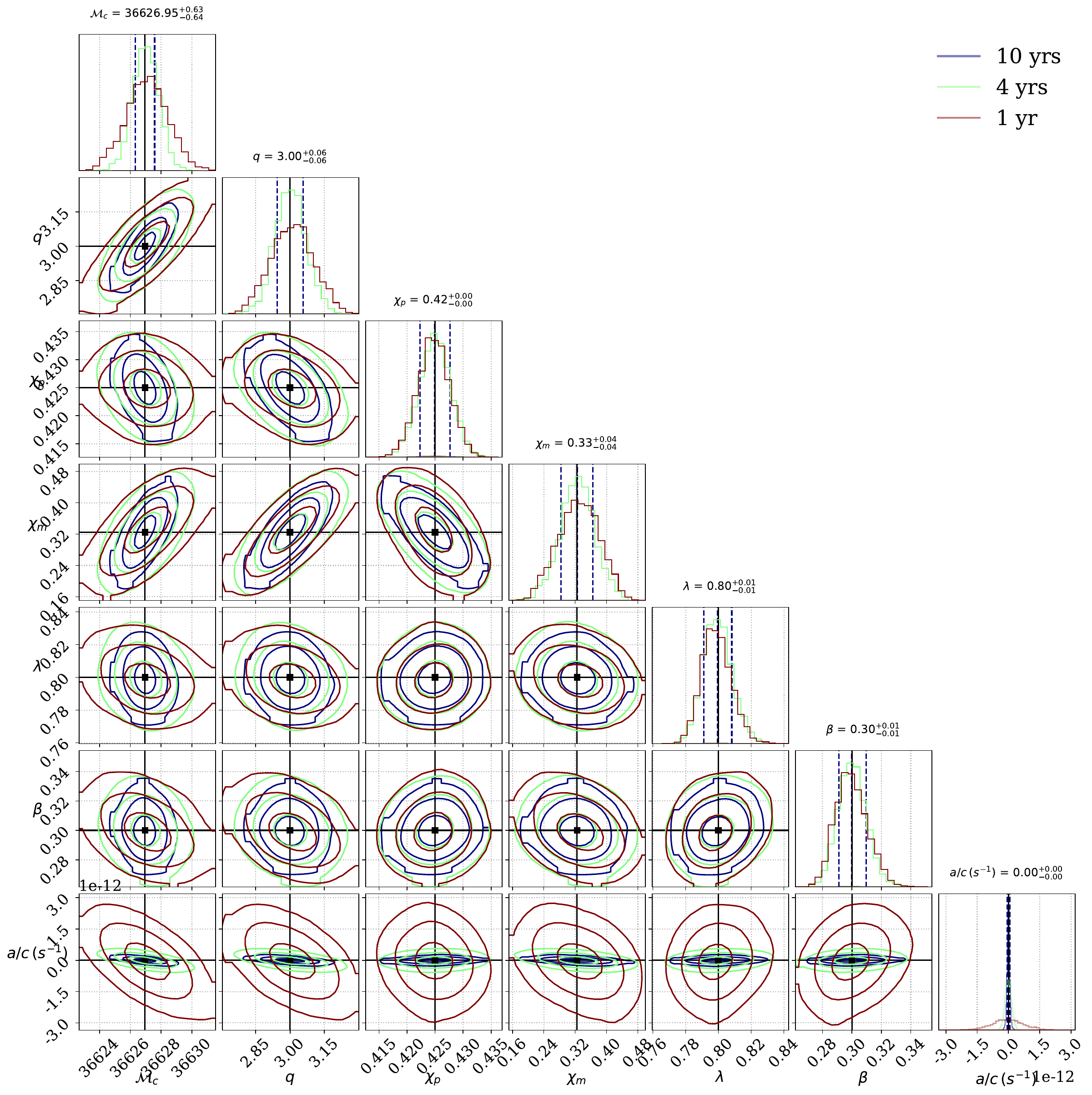} 
    \caption{Corner plot of the parameters used in the simulations underlying Fig.~\ref{fig:acc_single} as a function of the observation duration.
    }
    \label{fig:fullcorner}
\end{figure*}

Fig.~\ref{fig:fullcorner} presents a full corner plot to show the effect of the length of observations on other parameters. The simulations underlying Fig.~\ref{fig:fullcorner} are based on the same source parameters used to produce the events in Fig.~\ref{fig:acc_single}. Fig.~\ref{fig:fullcorner} provides additional information that is not shown in Fig.~\ref{fig:acc_single}. It shows that extending the observation duration is not only beneficial for improving the accuracy of acceleration measurement, but is also crucial for other parameters, especially for chirp mass and mass ratio.

\section{Towards full PN expressions of phase difference considering acceleration and jerk}\label{app:full_expression}

In the main text, we derived the LOSA and jerk-induced gravitational-wave phase differences at 1PN order. In this appendix, we present how these results could be generalized straightforwardly to the full 3.5PN order with spin contributions, while avoiding giving these long expressions explicitly.

As an illustration, we consider the full 3.5PN time-frequency relation for the nonspinning sector, which is given by \cite{Blanchet:2013haa}:
\begin{widetext}
\begin{multline}
\begin{split}
t_{NS}(f) = & -\frac{5M}{256\eta} v^{-8} \Bigg\{ 1 + \left( \frac{743}{252} + \frac{11}{3}\eta \right) v^2 \\
& - \frac{32\pi}{5} v^3 + 2\left( \frac{3058673}{1016064} + \frac{5429}{1008}\eta + \frac{617}{144}\eta^2 \right) v^4 - \left( \frac{7729}{252} - \frac{13}{3}\eta \right) \pi v^5 \\
& + \Bigg[ -\frac{10052469856691}{23471078400} + \frac{128}{3}\pi^2 + \frac{6848}{105}\gamma \\
& \quad + \left( \left( \frac{3147553127}{3048192} - \frac{451}{12}\pi^2 \right)\eta - \frac{15211}{1728}\eta^2 + \frac{25565}{1296}\eta^3 + \frac{6848}{105}\ln(4x) \right) v^6 \\
& - \left( \frac{15419335}{127008} + \frac{75703}{756}\eta - \frac{14809}{378}\eta^2 \right) \pi v^7 + \frac{65\Gamma M}{1536\eta} v^{-8} \\
& \Bigg\{ 1 + \frac{22}{13}\left( \frac{743}{252} + \frac{11}{3}\eta \right) v^2 - \frac{128}{13}\pi v^3 + \left( \frac{1755623}{122304} + \frac{32633}{1092}\eta + \frac{1101}{52}\eta^2 \right) v^4 \\
& \quad - \left( \frac{83228}{1365} + \frac{4592}{195}\eta \right) \pi v^5 \\
& + \Bigg[ -\frac{2274117187691}{5029516800} + \frac{66304}{975}\pi^2 + \frac{13696}{195}\gamma + \left( \frac{3311653861}{2830464} - \frac{3157}{78}\pi^2 \right)\eta \\
& \quad + \frac{2030687}{33696}\eta^2 + \frac{35693}{648}\eta^3 + \frac{13696}{195}\ln(4x) \Bigg] v^6 \\
& \quad - \left( \frac{158992529}{687960} + \frac{1015907}{4095}\eta - \frac{838}{4095}\eta^2 \right) \pi v^7 \Bigg\} \Bigg\} ,
\label{tofNS35PN}
\end{split}
\end{multline}
\end{widetext}
where NS indicates excluding all spin-dependent terms.
Eq.~\eqref{delayetotal} shows the arrival-time difference induced by the line-of-sight motion of the binary.
Substituting $t_{NS}(f)$ from Eq.~\eqref{tofNS35PN} into Eq.~\eqref{delayetotal}, we obtain the gravitational-wave phase difference induced by the line-of-sight acceleration as
\begin{equation}
    \Delta\psi_{NS}(a)
    =2\pi f\,\frac{1}{2}\Gamma t_f^2
    =\pi f \Gamma\,t_{NS}^{2}(f)\,,
    \label{dpsi_a_NS}
\end{equation}
while the phase difference induced by the line-of-sight jerk is
\begin{equation}
    \Delta\psi_{NS}(j)
    =2\pi f\,\frac{\frac{1}{6}j^{\parallel}t_f^3}{c}
    =\frac{\pi f j^{\parallel}}{3c}\,t_{NS}^{3}(f)\,.
    \label{dpsi_j_NS}
\end{equation}
Eq.~\eqref{dpsi_a_NS} and \eqref{dpsi_j_NS} are the compact nonspinning results for the LOSA- and jerk-induced phase differences. Re-expanding these relations in PN fashion in the small parameter $v$ would yield the dephasing at the desired order readily, up to the starting 3.5PN order. After re-expansion, the phase correction expression is checked to agree with the 3.5PN result in \citet{Lazarow:2024detectingaccelerating, Vijaykumar:2023tjg}.
The same procedure can readily be used to compute the dephasing contributions with aligned spins, simply by including these terms in $t_f$. The full term-by-term $3.5$PN expressions are available upon request.

\section{Comparisons between the multiple PN expansion levels and full PN structure in LOSA terms}
\label{app:-4PNvsfullPN}

Here we compare the next-to-leading PN-order effect of LOSA, in the form of its $-$3PN term used throughout our study, against the higher PN-order expression at the $3.5$ PN relative order, that is to say, including all negative PN orders up to $-$0.5PN provided by the application of the LOSA deviation to higher-order PN terms as in App.~\ref{app:full_expression}.
For example, the LOSA perturbation applied to the 1PN correction results in a $-3$PN term (1PN minus 4PN).
This is a higher-order term with respect to the leading LOSA $-$4PN correction as it is expanded in both the LOSA parameter and the usual PN parameter (and thus is smaller than both the leading 0PN standard term and the leading -4PN LOSA term).

\begin{table*}[htbp]
\centering
\caption{Phase differences of LOSA phase corrections at different PN orders relative to the preceding PN order, at a frequency corresponding to 10 years pre-merger, together with the \(\Gamma\) values for different total masses taken to be at the level of the 5-sigma constraints.}

\label{tab:pn_phase_diff_10yr}
\renewcommand{\arraystretch}{1.35}
\begin{tabular*}{\textwidth}{@{\extracolsep{\fill}}lccccccc@{}}
\toprule
\multirow{2}{*}{Total mass ($M_{\rm \odot}$)} & \multirow{2}{*}{\(\Gamma=5 \sigma_{\Gamma}(s^{-1})\)} & \multicolumn{6}{c}{Phase difference $\Delta\Psi_{\rm 0PN}-\Delta\Psi_{\rm 1 PN}$, $\Delta\Psi_{\rm \frac{n-1}{2} PN}-\Delta\Psi_{\rm \frac{n}{2} PN}$ [rad], (n=3, 4, 5, 6, 7)} \\
\cmidrule(lr){3-8}
 & & 1PN & 1.5PN & 2PN & 2.5PN & 3PN & 3.5PN \\
\midrule
$10^{4}\,M_\odot$
 & $2.145\times10^{-13}$ & $5.702\times10^{-1}$ & $-1.413\times10^{-1}$ & $4.708\times10^{-3}$ & $-2.354\times10^{-3}$ & $1.734\times10^{-4}$ & $-1.247\times10^{-4}$ \\
$10^{5}\,M_\odot$
 & $4.58\times10^{-13}$& $5.135\times10^{-1}$ & $-1.697\times10^{-1}$ & $7.539\times10^{-3}$ & $-5.029\times10^{-3}$ & $5.284\times10^{-4}$ & $-4.374\times10^{-4}$ \\
$10^{6}\,M_\odot$
 &$3.350\times10^{-12}$ & $1.584$ & $-6.979\times10^{-1}$ & $4.135\times10^{-2}$ & $-3.679\times10^{-2}$ & $5.484\times10^{-3}$ & $-3.292\times10^{-3}$ \\
$10^{7}\,M_\odot$
 & $1.985\times10^{-10}$ & $15.574$ & $-9.935$ & $8.296\times10^{-1}$ & $-1.106$ & $2.579\times10^{-1}$ & $-1.976\times10^{-1}$ \\
\bottomrule
\end{tabular*}
\end{table*}

\begin{table*}[htbp]
\centering
\caption{Phase differences of LOSA phase corrections at different PN orders relative to the preceding PN order, at a frequency corresponding to an accumulated $\mathrm{SNR}=1$, together with the \(\Gamma\) values for different total masses taken to be at the level of the 5-sigma constraints.}

\label{tab:pn_phase_diff_snr}
\renewcommand{\arraystretch}{1.35}
\begin{tabular*}{\textwidth}{@{\extracolsep{\fill}}lccccccc@{}}
\toprule
\multirow{2}{*}{Total mass $M_{\rm tot}$} & \multirow{2}{*}{\(\Gamma=5 \sigma_{\Gamma}(s^{-1})\)} & \multicolumn{6}{c}{Phase difference $\Delta\Psi_{\rm 0PN}-\Delta\Psi_{\rm 1 PN}$, $\Delta\Psi_{\rm \frac{n-1}{2} PN}-\Delta\Psi_{\rm \frac{n}{2} PN}$ [rad], (n=3, 4, 5, 6, 7)} \\
\cmidrule(lr){3-8}
 & & 1PN & 1.5PN & 2PN & 2.5PN & 3PN & 3.5PN \\
\midrule
$10^{4}\,M_\odot$
 & $2.145\times10^{-13}$ & $4.472\times10^{-1}$ & $-1.133\times10^{-1}$ & $3.859\times10^{-3}$ & $-1.973\times10^{-3}$ & $1.493\times10^{-4}$ & $-1.231\times10^{-4}$ \\
$10^{5}\,M_\odot$
 & $4.58\times10^{-13}$ & $3.312\times10^{-1}$ & $-1.139\times10^{-1}$ & $5.267\times10^{-3}$ & $-3.656\times10^{-3}$ & $4.033\times10^{-4}$ & $-3.747\times10^{-4}$ \\
$10^{6}\,M_\odot$
 &$3.350\times10^{-12}$& $2.150\times10^{-1}$ & $-1.136\times10^{-1}$ & $8.070\times10^{-3}$ & $-8.608\times10^{-3}$ & $1.597\times10^{-3}$ & $-1.225\times10^{-3}$ \\
$10^{7}\,M_\odot$
 & $1.985\times10^{-10}$ & $5.566\times10^{-1}$ & $-4.819\times10^{-1}$ & $5.610\times10^{-2}$ & $-9.800\times10^{-2}$ & $3.270\times10^{-2}$ & $-2.758\times10^{-2}$ \\
\bottomrule
\end{tabular*}
\end{table*}

To verify the impact of different PN orders on the accumulated phase, we present the phase differences relative to the 3.5PN order under two frequency cutoff scenarios (Tables~\ref{tab:pn_phase_diff_10yr} and \ref{tab:pn_phase_diff_snr}). The first scenario applies a frequency cutoff corresponding to exactly 10 years before the merger. The results indicate that the phase differences introduced by PN orders beyond \(-4\)PN and \(-3\)PN are generally small, suggesting that the leading and next-to-leading LOSA terms suffice for accurate waveform modeling. 

Although the 10-year cutoff yields seemingly large phase differences between the \(-4\)PN and \(-3\)PN terms for total masses of \(10^6 M_\odot\) and \(10^7 M_\odot\), it is important to note that during the early inspiral, the signal-to-noise ratio (SNR) of these heavy systems is extremely low. The 10-year integration therefore includes a vast number of cycles that are practically undetectable. To account for this observational limitation, the second scenario adopts a frequency cutoff defined by the epoch when the SNR reaches 1, meaning that we only accumulate the phase after the signal becomes detectable. Under this realistic SNR=1 cutoff, the phase differences between the \(-4\)PN and \(-3\)PN terms become negligible even for the heaviest systems. This confirms that truncating the PN expansion at the \(-4\)PN and \(-3\)PN level is well justified for practical detection purposes.

\begin{figure}[!htb] 
    \centering
    \includegraphics[width=0.5\textwidth]{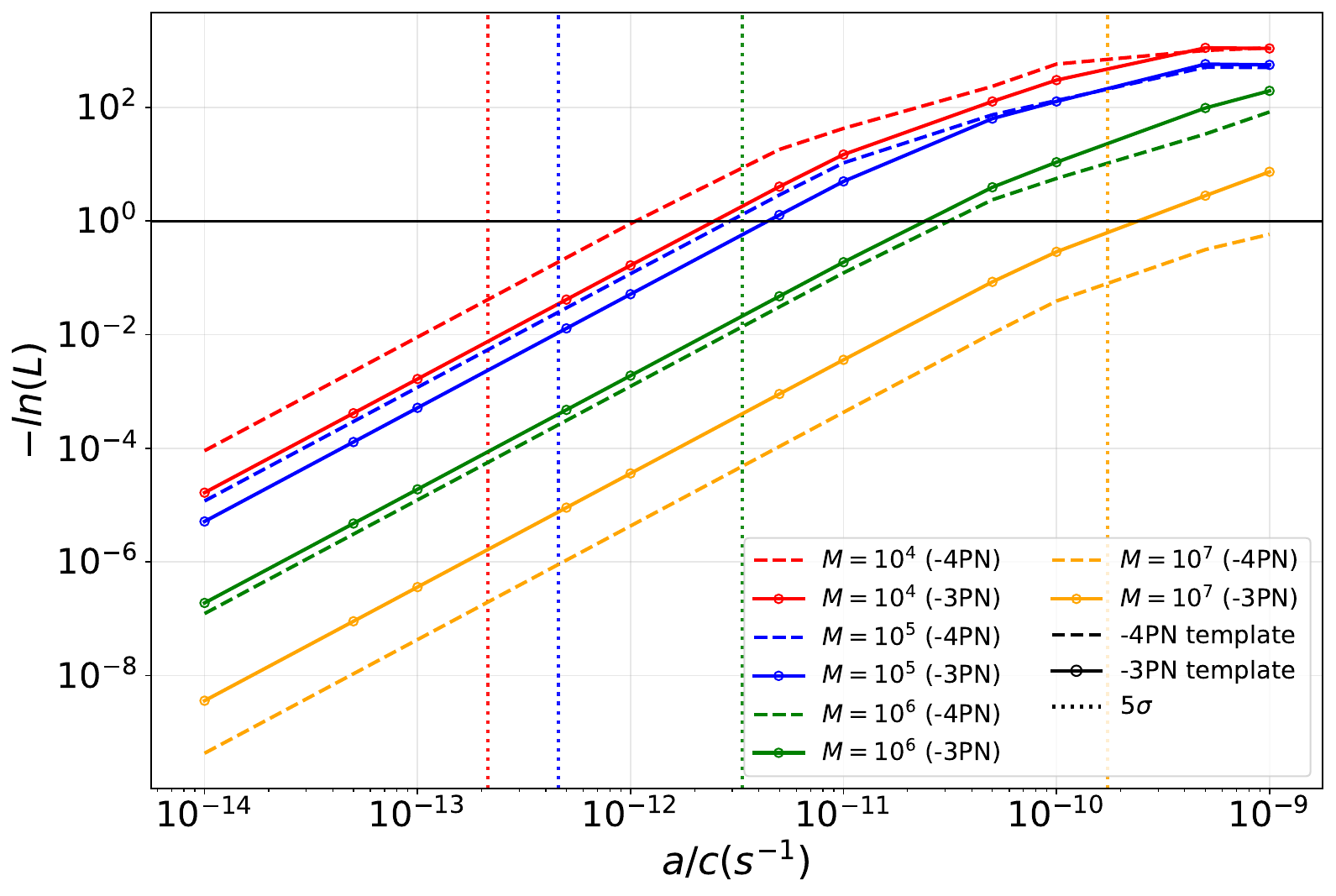} 
    \caption{Negative log-likelihood as a function of $ a/c$. The likelihood is computed with two different template waveforms to be compared with the full PN waveform, which retain only the leading-order LOSA phase correction term at $-$4PN (dashed lines) and up to $-$3PN (solid lines with open circles), respectively. For each template, the color encodes the total mass of the binary. The dotted vertical lines mark the $5\sigma$ sensitivity corresponding to each mass.
    }
    \label{fig:fullPN}
\end{figure}

Figure~\ref{fig:fullPN} reports the negative log-likelihood
between a simulated GW signal generated using the full expansion of LOSA terms from $-$4PN to $-$0.5PN,
and two template waveforms: one retaining only the leading-order term at $-$4PN (dashed lines),
and one including terms up to $-$3PN (solid lines with open circles).
The color of each curve encodes the total mass of the binary, and the dotted vertical lines mark the $5\sigma$
constraints on $a/c$ for the corresponding masses. The figure shows that, for both templates, the likelihood penalty remains
modest over the entire range of $a/c$ probed here, and in particular it stays below unity at all the
$5\sigma$ thresholds. This indicates that, although the PN template is not a perfect match
to the full PN signal, the residual mismatch is small enough that it does not compromise detection:
the next-to-leading-order LOSA approximation at $-$3PN is sufficient for all our purposes.

\section{Breakdown of Fisher Matrix Parameter Estimation at Low Frequencies}
\label{app:fisher_low_freq}

\begin{figure*}
    \centering
    \includegraphics[width=\textwidth]{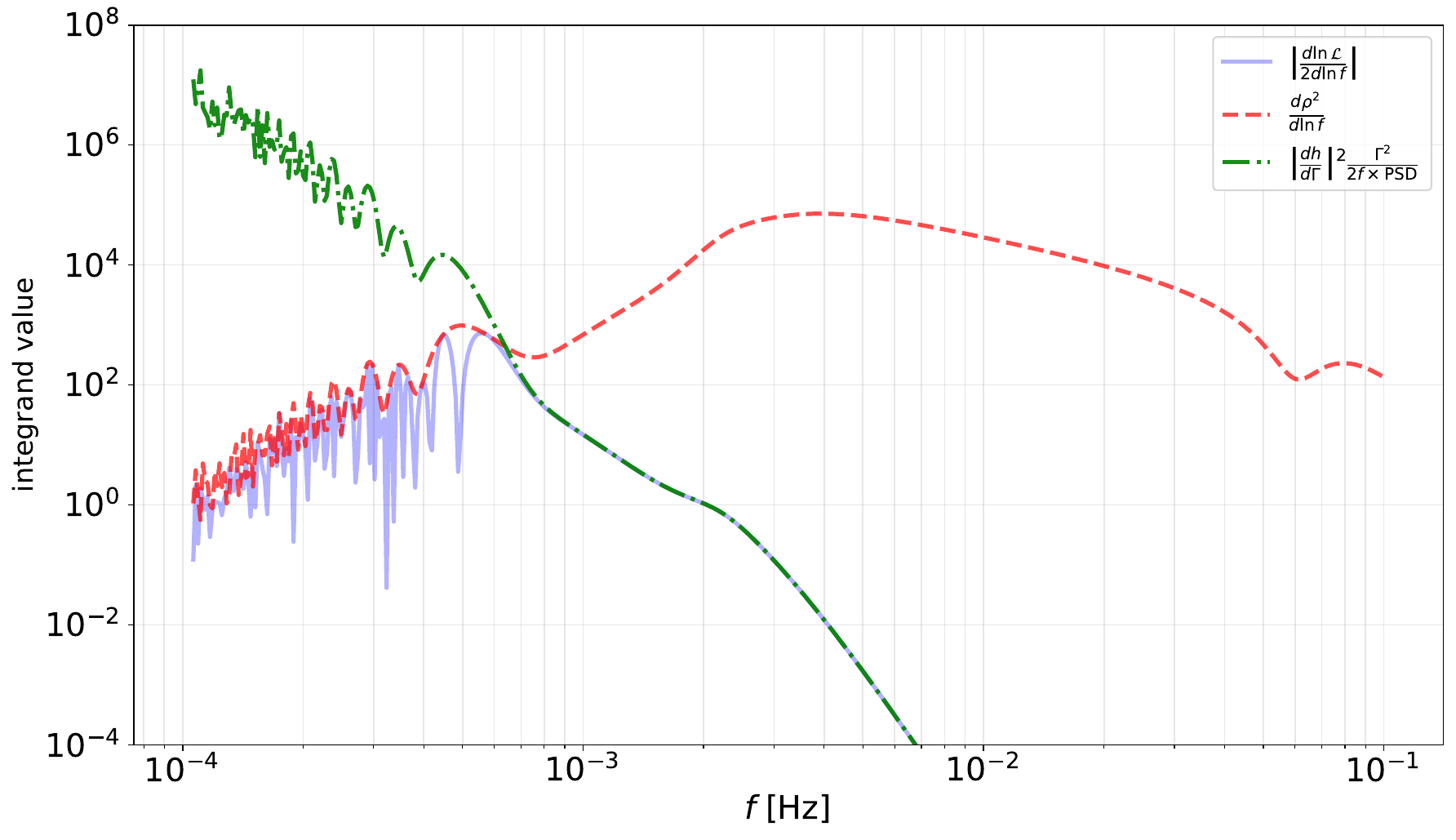} 
\caption{
Comparison between the likelihood integrand (blue solid), the Fisher matrix integrand (green dash-dotted), and the SNR integrand (red dashed) as a function of frequency, for an MBHB with total mass $M = 10^5 \, M_\odot$ (other parameters are shown in Table \ref{other paras} in Appendix \ref{appendix paras}), with a line-of-sight acceleration $ a= 0.01 \, {\rm m/s^2}$. The likelihood here compares $a=0$ to $a\neq 0$.
The blue curve shows $|d \ln \mathcal{L} / (2 d \ln f)|$, representing the exact likelihood integrand; the green curve shows $|d h / d \Gamma|^2 \, \Gamma^2 / (2 f \times {\rm PSD})$, representing the integrand of the likelihood in the quadratic approximation at the heart of the Fisher matrix approach.  
The red dashed curve shows $d {\rho}^2 / d \ln f$, serving as an upper bound: for pure dephasing, the likelihood is bounded by the SNR at all frequencies as shown by Eq.~\eqref{eq:snr_likelihood_density}. 
In the low-frequency regime ($f \lesssim 10^{-3} \, {\rm Hz}$), the green curve diverges above both the blue and red curves, indicating that the Fisher matrix approximation breaks down and significantly overestimates the available information: the large acceleration-induced phase difference should be mitigated by the small amplitude.
}
\label{fig:whyfishernot}
\end{figure*}

This appendix analyzes the validity of the Fisher matrix approximation in the low-frequency regime for investigating the measurability of the LOSA effect.
The Fisher information matrix $\Gamma_{ij}$ and its inverse give a Gaussian estimate of parameter estimation errors (see the main text and Eqs.~\eqref{eq:snr}, \eqref{eq:likelihood} for the definitions of the Fisher matrix, the SNR, and the likelihood function in Sec.~\ref{data}).

The key issue is that the derivation of the Fisher matrix makes use of the linearized signal approximation, writing that, around the true parameters,
\begin{equation}
    s(\theta) = s(\theta_0) + \partial_i s \Delta \theta^i + \mathcal{O}(\Delta \theta^2) \,,
\end{equation}
where $\Delta \theta = \theta - \theta_0$ is the vector of parameter deviations. For a signal decomposed as an amplitude and phase, $s(f) = A(f) \exp[-i \Psi(f)]$, its derivative with respect to the component $i$ of the parameter vector takes the form
\begin{equation}
    \partial_i s(f) = [\partial_i A / A - i \partial_i \Psi](f) A(f) e^{-i \Psi(f)} \,.
\end{equation}
In the specific case of LOSA, the dephasing induced by the acceleration $\Delta \Psi$ is very steeply increasing at early times / low frequencies, where the amplitude is small (compared to the noise PSD). In this region, the phase derivative term $\partial_{\Gamma}$ ends up saturating the linearized signal approximation and giving a spuriously large contribution to the Fisher matrix element. More concretely, we can visualize what contributes most to the Fisher matrix element as follows: consider a 1D parameter estimation in just the $\Gamma$ parameter, keeping all other parameters fixed; inject a value of 0, and compute the log-likelihood for non-zero values. As in Eq.~\eqref{eq:snr_likelihood_density}, neglecting amplitude and response effects, the density of the log-likelihood is
\begin{equation}
    \frac{d\ln\mathcal{L}}{d\ln f}
\simeq -\frac{1}{2} |1-e^{-i\Delta\Psi (\Gamma)}|^{2}\,\frac{d\rho^{2}}{d\ln f},
\end{equation}
while applying the linearized-signal approximation, the Gaussian approximation is equivalent to using the following quadratic for this log-likelihood density:
\begin{equation}
    \left. \frac{d\ln\mathcal{L}}{d\ln f} \right|_{\mathrm{Fisher}}
\simeq -\frac{1}{2} \left( \partial_{\Gamma} \Delta\Psi \right)^{2} \Gamma^2\,\frac{d\rho^{2}}{d\ln f},
\end{equation}
It is then possible for the prefactor to exceed the bound of 2, giving a greater likelihood penalty than would be allowed by the SNR density, and therefore spurious extra information.

We illustrate the issue in Fig.~\ref{fig:whyfishernot}, using as an example an MBHB system with $M = 10^5 \, M_\odot$, $q = 3$, $\rm a = 0.01 \, {\rm m/s^2}$. In the low-frequency band ($f \lesssim 10^{-3} \, {\rm Hz}$), the green curve (Fisher-matrix-approximated log-likelihood density) rises above the red curve (SNR density), violating the theoretical expectation. The divergence of the acceleration-induced phase causes the Fisher matrix to rapidly accumulate unphysical ``information'' as we lower the starting frequency by increasing the total observation time, giving overoptimistic constraints on $ \sigma_{\rm {a}/{c}}$.

These observations constitute the main reason why we chose to perform only MCMC parameter estimation to produce all results in our study.
We point out that the breakdown of the Fisher matrix approximation that we uncovered in our investigation is not directly linked to a numerical instability of the signal derivatives, and would not be visible by adjusting the steps used to compute numerical derivatives; as such, it could potentially affect previously published results in the literature,
in the absence of comparisons between Fisher and full parameter estimation results. We also note that the same kind of issue will affect tests looking for negative-PN effects: environmental effects such as accretion and dynamical friction, and, to a lesser extent, the $-$1PN dephasing characteristic of dipolar radiation in the context of tests of General Relativity.

%\clearpage  
\bibliographystyle{apsrev4-2}
\bibliography{biblio}

\end{document}